\documentclass[prb,floatfix,superscriptaddress,twocolumn,aps,showpacs,amsmath,amssymb]{revtex4}
\tighten
\renewcommand{\narrowtext} 
{\begin{multicols}{2}\global\columnwidth20.5pc} 
\newcommand{\be}{\begin{equation}}
\newcommand{\ee}{\end{equation}}
\newcommand{\bea}{\begin{eqnarray}}
\newcommand{\eea}{\end{eqnarray}}

\newcommand{\ci}{\mathfrak{i}}

\usepackage{wasysym}
\usepackage{graphicx}
\usepackage{bm}

\usepackage{ulem}
\usepackage{color}
\definecolor{darkred}{rgb}{0.6,0,0}
\definecolor{darkblue}{rgb}{0.0,0,0.6}
\definecolor{red}{rgb}{1,0,0}

\usepackage{multirow}
\usepackage{textcomp}

\begin{document} 
%\preprint{p29.C60Long}

\title{Transport properties of individual C$_{60}$-molecules}

\author{G. G\'eranton}
\affiliation{ Institute of Nanotechnology,
 Karlsruhe Institute of Technology, Campus North, D-76344 Karlsruhe, Germany}
\author{C.  Seiler}
\affiliation{ Institute of Nanotechnology,
 Karlsruhe Institute of Technology, Campus North, D-76344
  Karlsruhe, Germany}
\affiliation{Center of Functional Nanostructures, 
 Karlsruhe Institute of Technology, Campus South, 
  D-76131 Karlsruhe, Germany}
\affiliation{Institut f\"ur Theorie der Kondensierten Materie,
 Karlsruhe Institute of Technology, Campus South, D-76128 Karlsruhe, Germany}
\author{A. Bagrets}
\affiliation{ Institute of Nanotechnology,
 Karlsruhe Institute of Technology, Campus North, D-76344
  Karlsruhe, Germany}
\affiliation{Steinbuch Center for Supercomputing, 
 Karlsruhe Institute of Technology, 
  D-76344 Karlsruhe, Germany}
\author{L. Venkataraman}
\affiliation{Department of Applied Physics and Applied Mathematics, Columbia University, New York, NY 10027}
\author{F. Evers}
\affiliation{ Institute of Nanotechnology,
 Karlsruhe Institute of Technology, Campus North, D-76344
  Karlsruhe, Germany}
\affiliation{Center of Functional Nanostructures, 
 Karlsruhe Institute of Technology, Campus South, 
  D-76131 Karlsruhe, Germany}
\affiliation{Institut f\"ur Theorie der Kondensierten Materie,
 Karlsruhe Institute of Technology, Campus South, D-76128 Karlsruhe, Germany}

\date{\today}% It is always \today, today,

%Use showkeys class option if keyword
\begin{abstract}
Electrical and thermal 
transport properties of C$_{60}$ molecules are investigated with density-functional-theory based calculations. 
These calculations suggest that the optimum contact geometry for an electrode terminated with a single-Au atom is through binding
to  one or two C-atoms of C$_{60}$ with a tendency to promote the sp$^2$-hybridization into an sp$^3$-type one.
Transport in these junctions is primarily through an unoccupied molecular orbital that is partly hybridized with the Au, which results in splitting the 
degeneracy of the lowest unoccupied molecular orbital triplet. The
transmission through these junctions, however, cannot be modeled by a
single Lorentzian resonance, as our results show evidence of quantum
interference between an occupied and an unoccupied orbital. The interference 
results in a suppression of conductance around the Fermi energy. 
Our numerical findings are readily analyzed analytically within  a simple two-level model. 
%Due to cancellation
%effects, a current induced by temperature gradients is left unaffected
%by this effect. 
\end{abstract}
%\pacs{PACS numbers: }
%\narrowtext
\maketitle

\section{Introduction.}
%The general transport properties of C$_{60}$-molecules constitute one of the longer standing topics in {\it Molecular Electronics}.
%%
C$_{60}$ on metal surfaces is an important model system for understanding basic processes
in binding of (conjugated) organic molecules to metal electrodes, and
has been studied in the past both experimentally and
theoretically. \cite{lu03, lu04, rogero02, sau08, torrente08, tamai08,
  abad10,hamada11} Transport through C$_{60}$ has motivated
 investigations by  experimentalists and theorists 
from early on with an emphasis on correlation physics 
like the Kondo effect \cite{park00, yu04, pasupathy04, parks07,roch08}
or vibrational degrees of freedom \cite{schulze08njp, schulze08prl,frederiksen08,gagliardi08,fock11}. 

%% copper surface
Concerning the linear conductance,
the situation seems to be particularly well investigated
with Cu-electrodes, where a combination of
ab-initio based calculations and STM-experiments have provided
a detailed understanding. \cite{neel07,schull09,schull10}
These investigations suggest that the conductance, $G$,
of C$_{60}$  bound to  Cu(111)-substrates
 is sensitive to the anchoring mechanism.
In general, $G$ is relatively large 
for a single Cu-atom contacting C$_{60}$ immobilized 
on a Cu(111) surface ($\sim$0.13 $\text{G}_0$, where
$\text{G}_0=2e^2/h$). \cite{schull11}
In contrast, the conductance of  C$_{60}$ on Au-electrodes
has not been studied extensively.
%LV: Fabian Pauly here??
Mechanical break junction experiments 
have found that the conductance of a junction can be as high as
$0.1-0.2$ G$_0$ \cite{boehler07,kiguchi08}
while STM-break junction experiments report much smaller values
with a very broad scattering in conductance histograms 
for Au-, Pt- and Ag-electrodes \cite{yee11}. 
%LV COMMENT: add Segalman reference here.

Understanding the transport mechanism in C$_{60}$/Au junctions is of
interest for two reasons.
First, any analysis and design of transport processes through
single molecules relies upon an understanding of the influence
of the electrodes. 
\cite{mayor03}
Second,  the specific molecule C$_{60}$-molecule might play a special
role in the context of contact formation  because 
it was recently proposed to be a suitable general  
anchor group  due to its size and electronic conjugation. \cite{martin08, leary11} 

%%%%%%%%%%%%%%%%%%%%%%%%%%%%%%%%%%%%%%%%%%%%%%%%%%%%%%%%%%%%%%%%%%%%%%%%%
%%%%%%%%%%%%%%%%%%%%%%%%%%%%%%%%%%%%%%%%%%%%%%%%%%%%%%%%%%%%%%%%%%%%%%%%%
%%%%%%%%%%%%%%%%%%%%%%%%%%%%%%%%%%%%%%%%%%%%%%%%%%%%%%%%%%%%%%%%%%%%%%%%%
In this paper we address the transport characteristics of single-molecule junctions formed using C$_{60}$ molecules attached to Au electrodes.
Our calculations are employing the DFT-based non-equilibrium Green's functions
(NEGF) formalism\cite{TranspForm}. Our first theoretical result is that Au-electrodes are invasive:
Au single adatoms have a tendency to form chemical bonds with C-atoms and thus
locally affect the sp$^2$-conjugation in C$_{60}$, similar to what
has been reported for Cu electrodes. 
We find that the alternative
scenario, where the adatom resides in a hexagonal/pentagonal facet
of C$_{60}$ is not energetically favored as it has a binding energy
that is 0.5eV lower. We thus confirm statements 
reported in Ref. \onlinecite{shukla08}. 
Our second result is that
C$_{60}$ has a slight tendency to charge negatively on Au 
with single adatom binding, though not as much as on
Cu (or Ag). \cite{shukla09}
This contrasts results on flat Au(111) where the charge transfer is negligible. 
\cite{sau08,torrente08}
As a consequence, with Au-electrodes transport is more
LUMO-dominated. 
However, in contrast with Cu, it will in general not be close to
resonant. 
%%%%%%%%%%%%%%%%%%%%%%%%%%%%%%%%%%%%%%%%%%%%%%%%%%%%%%
%% structure optimization
%%
%%
%%%%%%%%%%%%%%%%%%%%%%%%%%%%%%%%%%%%%%%%%%%%%%%%%%%%%%
\begin{figure*}[thp]
\begin{center}
\begin{tabular}{ccc}
\includegraphics[width=0.31\linewidth]{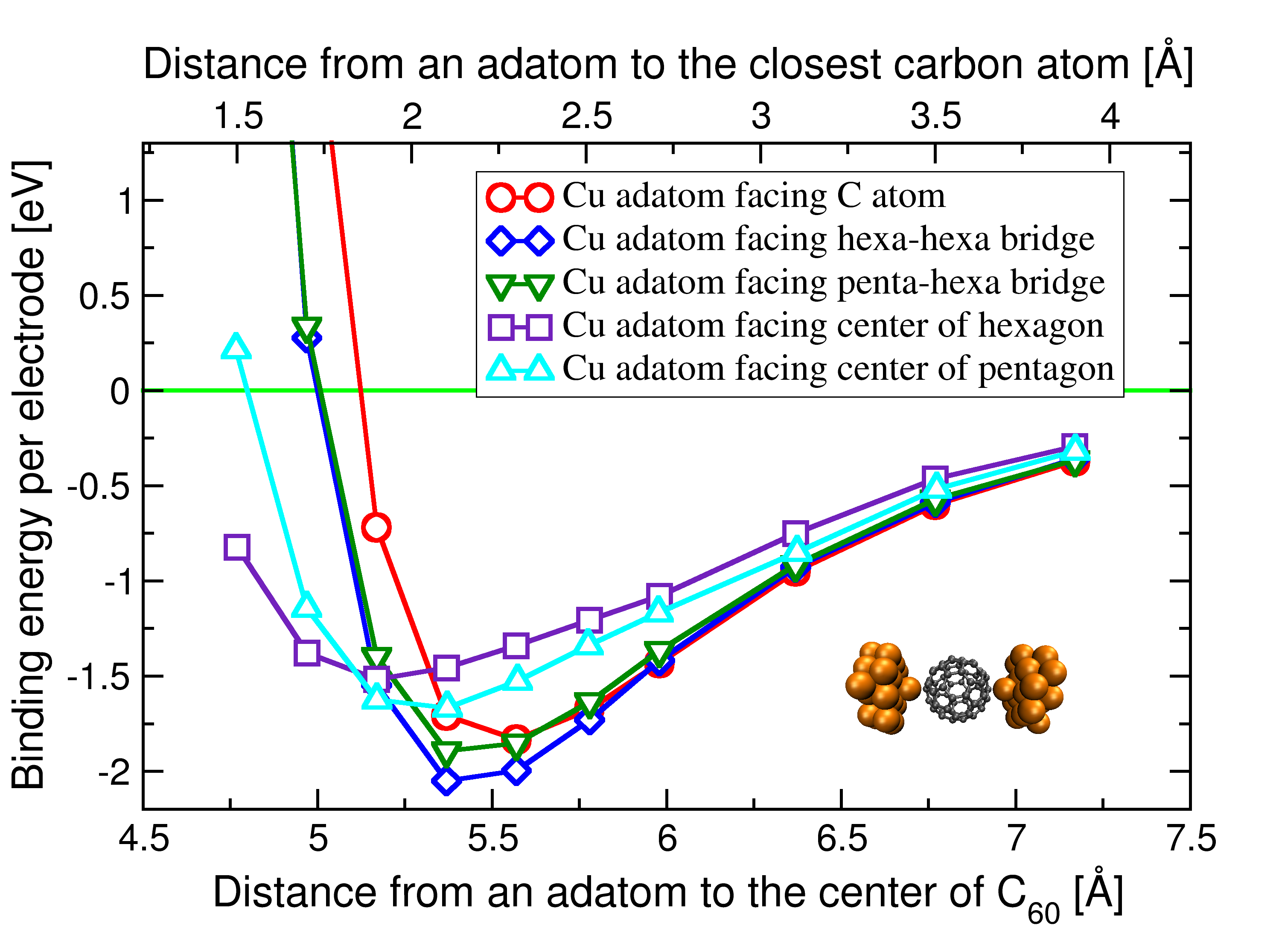} &
\includegraphics[width=0.31\linewidth]{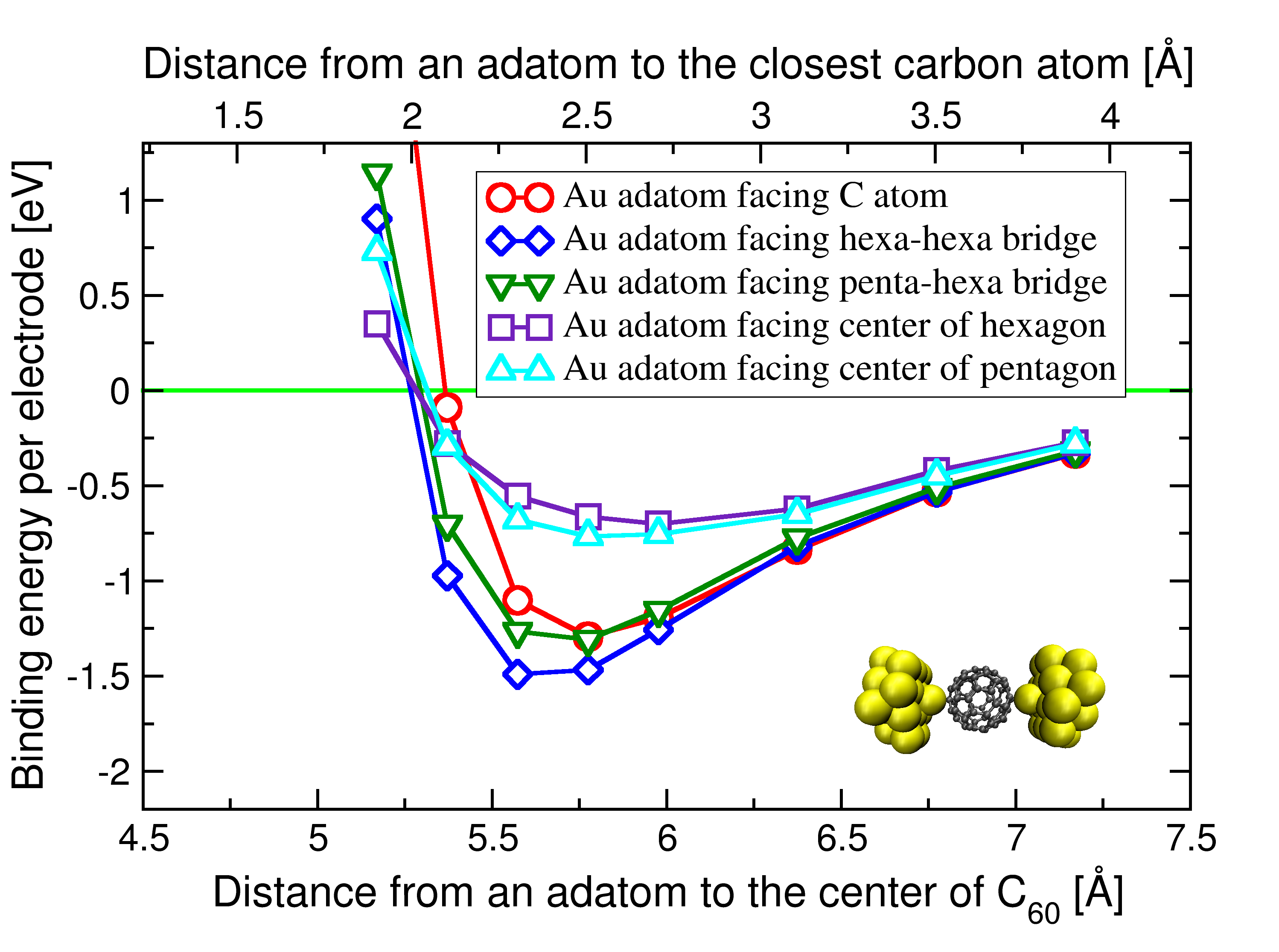} &
\includegraphics[width=0.31\linewidth]{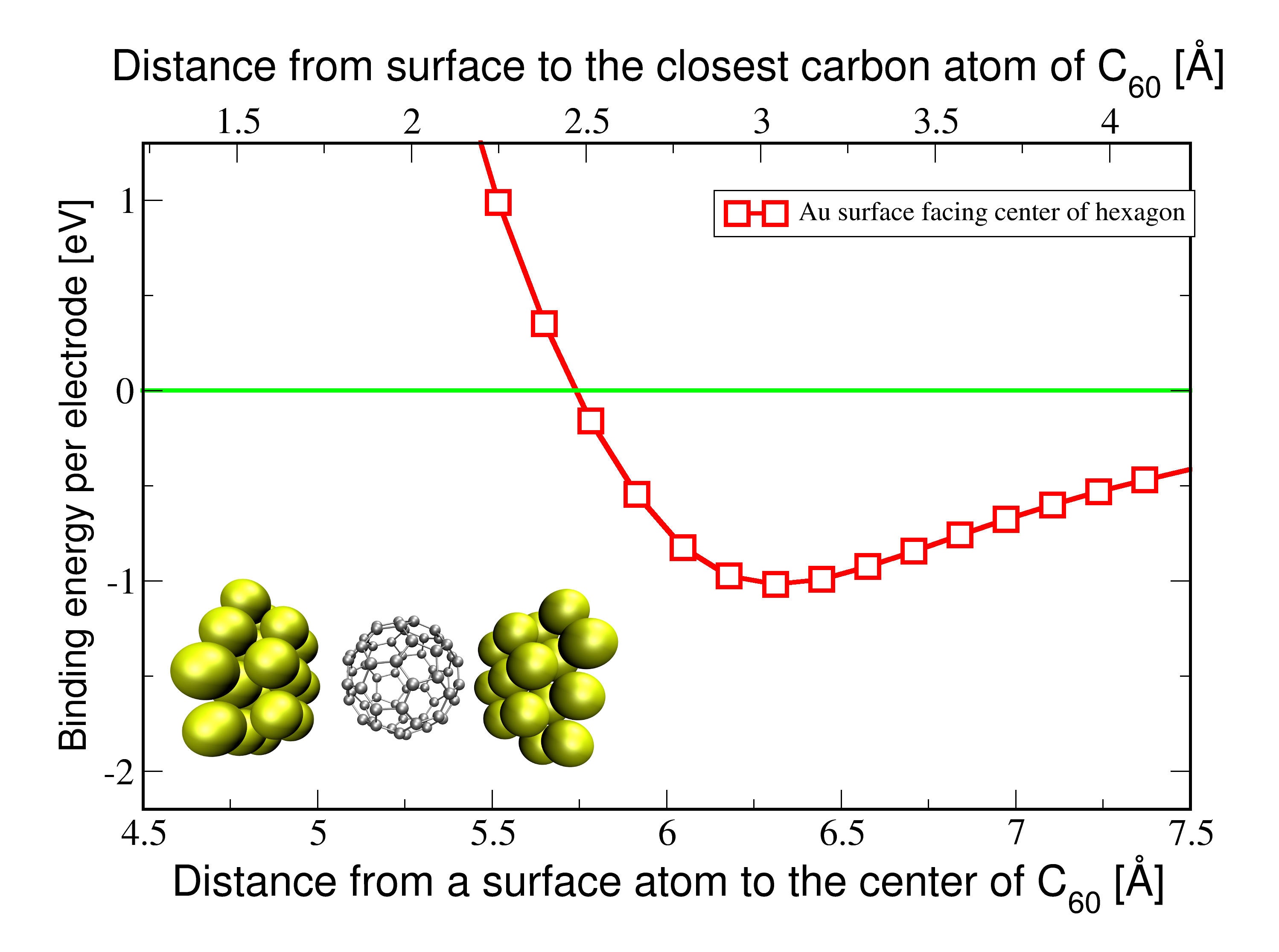} \\
(a) &
(b) &
(c)
\end{tabular}
\caption{Energy profiles from DFT with Grimme corrections:
  Adatom geometries with Cu-electrodes (a)
  and Au-electrodes (b).
  (c): Flat Au-electrodes facing
  C$_{60}$ hexagons.}
\end{center}
\label{f1}
\end{figure*}
%%%%%%%%%%%%%%%%%%%%%%%%%%%%%%%%%%%%%%%%%%%%%%%%%%%%%%
Third, we determine the transport characteristics of several
C$_{60}$-junctions that exhibit somewhat different contact
geometries by calculating the transmission function, $T(E)$, and the
Seebeck-coefficients, $S(E)$. 
For two geometries data similar to ours have 
been reported before\cite{bilan12}. Our analysis goes 
significantly beyond earlier work because 
we provide evidence that interference between transport channels 
plays a quantitatively important role, especially when
the Fermi-energy is situated between HOMO- and LUMO-resonances of the 
C$_{60}$ molecule.  We show that 
%%LV Why do you want to take this out??
deviating from  earlier claims\cite{ono07}
the conductance at the Fermi energy is not resonant but rather
strongly suppressed due to destructive interference from 
two strongly  coupled transport channels. This suppression
leads to a sharp, step-like increase of the
Seebeck-coefficient near the minimum transmission energy. 
We analyze our findings within an analytic model for a two-level system
and find that it supports the results of
DFT-based transport calculations.
 In addition, this analysis also shows
  that due to cancellation effects, electrical currents
  driven by heat gradients (rather than by a bias voltage)
  remain almost unaffected by destructive interference effects.

%%%%%%%%%%%%%%%%%%%%%%%%%%%%%%%%%%%%%%%%%%%%%%%%%%%%%%
\section{Symmetric contacts with single adatom}
%%%%%%%%%%%%%%%%%%%%%%%%%%%%%%%%%%%%%%%%%%%%%%%%%%%%%%

%%%%%%%%%%%%%%%%%%%%%%%%%%%%%%%%%%%%%%%%%%%%%%%%%%%%%%
\subsection{Method}
We compute the total energy of the extended molecule (C$_{60}$ plus
contact model)  for different contact geometries using
density functional theory (DFT). 
To this end, we employ the TURBOMOLE package \cite{turbomole89} with the 
BP86 functional\cite{becke88} and van der Waals interactions
included on the level of Grimme corrections  \cite{Grimme2004};
see appendix \ref{AppA} for details on them.
The geometry was optimized in the following way: the relative
position of all electrode atoms was fixed with bond-lengths as given
by Au-bulk. The geometry of C$_{60}$ was optimized in vacuum.
Thereafter, C$_{60}$ was approached to the electrode with different 
surface-molecule contact geometeries: 
hexagon or pentagon on  the flat surface or 
atop (C-Au-atom) and bridge position for single-adatom geomeries, 
see Fig. \ref{f1}. 
For each contact geometry a trace binding-energy  was recorded as a
function of distance.
In this process the geometry of C$_{60}$ was fixed. 

%%%%%%%%%%%%%%%%%%%%%%%%%%%%%%%%%%%%%%%%%%%%%%%%%%%%%%
\subsection{Results: Adatom geometries}
%%%%%%%%%%%%%%%%%%%%%%%%%%%%%%%%%%%%%%%%%%%%%%%%%%%%%%

Binding energies for different inter-electrode distances are displayed
in Fig.~\ref{f1}. 
One infers from Fig. \ref{f1}a,b that  Cu or Au-adatoms prefer bonding 
in bridge positions, where two hexagons touch each other
({\it hexa-hexa-bridges}).  However, for larger distances between the electrodes,
i.e. more generally for
relaxation under external constraints,  adatoms may also sit
on-top of C-atoms  or in {\it penta-hexa} bridge position.
 Since the energy difference between these three geometries is under
 100 meV  even close to the minima, these three geometries should be 
energetically accessible, especially in break-junction experiments
where  the electrode structures can deviate from pyramidal tips.
%%%  Since the energy cost for geometry transformation can be seen in Fig. \ref{f1} to fall 
%%% below 100meV even near minimum positions, all of these three geometries should be
%%% considered as relevant.  {\color{blue} This is especially true in the case of break junction
%%% experiments, since there the electrode geometry does not follow, in
%%% general, a simple geometry and therefore effects of residual
%%% strain should be a common encounter. }
%%%

%\subsection{Discussion}
{\bf Discussion:} Carbon atoms of C$_{60}$ in vacuum are $sp^2$-hybridized.
However, the bond angle (108$^\circ$) is relatively far from the 
``ideal''  $sp^2$-hybridization value
($\approx$ 120$^\circ$).  Therefore, it is plausible that C$_{60}$ is susceptible for
bonding with adatoms\cite{schneebeli11}, 
in contrast to simple $sp^2$-hybridized carbon
like graphene or other conjugate molecules.
Fig. \ref{f1}a,b reveal that the binding energy
of Cu to C$_{60}$ exceeds that of Au by 0.5eV.
On a qualitative level this observation
could be related
  to the fact that the KS-work function of Cu is situated considerably
  above the one of C$_{60}$ in vacuum (see Fig. \ref{FspecC60}).
  Therefore, a moderate flow of electrons into the C$_{60}$-LUMO
  might increase the interaction with the Cu-surface as compared to
  the one of Au.

The relatively strong interaction between the carbon
  $\pi$-system and the Au-adatom also manifests itself in the ratio
  of the GGA binding energy (ignoring Grimme the corrections)  
to the total binding energy, $\rho\equiv
E_{\text{GGA}}/(E_{\text{GGA}}+E_{\text{Grimme}})$.
It  helps to quantify how close 
the bond is to being covalent. If $\rho \approx 1$, the van der Waals 
contribution is negligible and a covalent chemical bond has formed. In contrast, if
$\rho\ll 1$, bonding is predominantly of the van der Waals
type and the adatom should be thought about as being physisorbed.

%%%%%%%%%%%%%%%%%%%%%%%%%%%%%%%%%%%%%%%%%%%%%%%%%%%%%%
\begin{table}[thbp]
\begin{center}
\begin{tabular}{|c|c|c|c|r|}
\hline
%Electrode 		& Position	& $l_\text{bond}$ [$\AA{}$] 	&  $E_\text{bind}$ [eV] 	& $E_\text{GGA}/E_\text{bind}$ \\ \hline
Electrode 		& Position	& $l_\text{bond}$ [\AA] 	&  $E_\text{bind}$ [eV] 	& $\rho$ \\ \hline
\multirow{3}{*}{Cu} 	& C-atom 	& 2.0  			& -1.83			&  66$\%$ \\\cline{2-5}
    			& p-h bridge  	& 2.2  			& -1.89			&  61$\%$ \\\cline{2-5}
    			& h-h bridge  	& 2.2  			& -2.05			&  63$\%$ \\ \hline 
\multirow{3}{*}{Au} 	& C-atom 	& 2.2  			& -1.29			&  56$\%$ \\\cline{2-5}
			& p-h bridge	& 2.4  			& -1.31			&  54$\%$ \\\cline{2-5}
			& h-h bridge	& 2.2  			& -1.49			&  54$\%$ \\ \hline 
Au (flat)		& hexagon 	& 3.0 			& -1.02			& -32$\%$ \\ \hline
\end{tabular}
\end{center}
\caption{Characteristics of lowest energy molecular junction
  geometries where $l_{\text{bond}}$ is the Au-C or Cu-C bond length,
  $E_{\text{bind}}$ the  binding energy per electrode and
  $\rho\equiv E_{\text{GGA}}/E_{\text{bind}}$ gives an indication
  of the covalent contribution to the bond.}
\label{TableAtomistic}
\end{table}
%%%%%%%%%%%%%%%%%%%%%%%%%%%%%%%%%%%%%%%%%%%%%%%%%%%%%%
The ratios $\rho$ are given in
Table~\ref{TableAtomistic} along with the corresponding total binding 
energies and bond lengths.
For both Cu and Au-electrodes, the  values $\rho \geq 1/2$ 
suggest that bonding is predominantly covalent consistent with the
expectations  formulated in the preceeding paragraph.

%%%%%%%%%%%%%%%%%%%%%%%%%%%%%%%%%%%%%%%%%%%%%%%%%%%%%%
%% Extension: flat electrodes
%%%%%%%%%%%%%%%%%%%%%%%%%%%%%%%%%%%%%%%%%%%%%%%%%%%%%%
\subsection{Results: Flat Au(111) electrodes}

We also consider flat Au(111) electrodes (without an
  adatom) facing a hexagon of C$_{60}$.
The binding energy profile of that configuration is displayed in
Fig.~\ref{f1}c. The bond distance here is measured from the position of
the nuclei of the first Au(111) layer. Most notably, the pure GGA
functional
without Grimme corrections would indicate that the configuration at a
distance of  $\sim 3.0\,\text{\AA}$ is
non-binding. This is reflected in the sign of $\rho$ in Table~\ref{TableAtomistic}. 
The bond has
no covalent contribution in this geometry, consistent with
the higher coordination
number of surface
atoms compared to adatoms. 
As shown in Table \ref{TableAtomistic}, 
the bond distances for the C$_{60}$-Au(111) geometry is $0.8-1.0\,\text{\AA}$ larger than in the case of C$_{60}$-adatom geometry.
The binding energy is of similar magnitude as in the adatom case, indicating that this configuration may
indeed be relevant in experiments.

%%%%%%%%%%%%%%%%%%%%%%%%%%%%%%%%%%%%%%%%%%%%%%%%%%%%%%
\section{Electronic Structure}
%%%%%%%%%%%%%%%%%%%%%%%%%%%%%%%%%%%%%%%%%%%%%%%%%%%%%% 

The nature of the molecule-electrode bond is of crucial importance for
C$_{60}$ transport properties. Since the formation of a covalent bond with C$_{60}$ implies
 that the conjugation of the $\pi$-electron system 
is broken at the contact, a transport barrier forms and the molecule
should be considered as ``weakly coupled''. 
In this section we discuss the effect of the bonding on the electronic
structure of the molecule.

%%%%%%%%%%%%%%%%%%%%%%%%%%%%%%%%%%%%%%%%%%%%%%%%%%%%%%
\begin{figure}[tbp]
\begin{center}
\includegraphics[width=0.5\columnwidth]{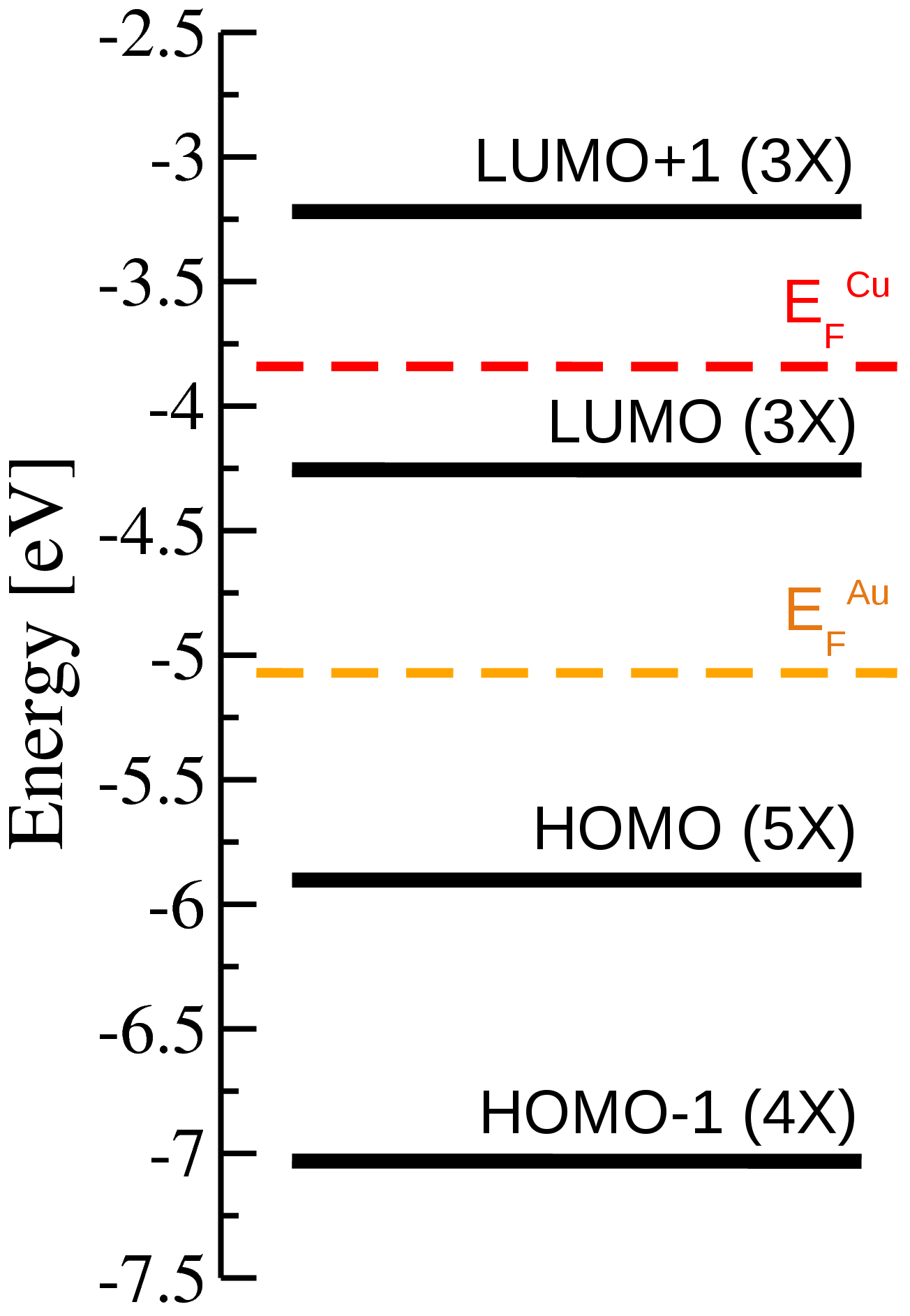}
\end{center}
\caption{Kohn-Sham energy levels of C$_{60}$ in
 vacuum (black lines).  Fermi energies of Cu (red) and Au (yellow)
estimated from  DFT calculation for the 24-atoms clusters 
used for the transport calculations.}
\label{FspecC60}
\end{figure}
%%%%%%%%%%%%%%%%%%%%%%%%%%%%%%%%%%%%%%%%%%%%%%%%%%%%%%
\subsection{Free molecule}
\label{sIIIA}
%% free molecule
The position of relevant molecular orbitals of C$_{60}$ in
vacuum, as obtained from DFT calculations, is shown in
Fig.~\ref{FspecC60}. The Highest Occupied Molecular Orbital (HOMO) and
Lowest Unoccupied Molecular Orbital (LUMO) levels are
5-fold and 3-fold degenerate. The associated HOMO-LUMO gap  is
$\Delta\sim 1.6$ eV, slightly underestimating the experimental
value 2.3 eV \cite{GapC60}. (For a discussion of this discrepancy, 
see Sec. \ref{ssIVE}.) 
%%LV
The position of the chemical potentials of
Au and Cu relative to C$_{60}$  energy levels are also shown in Fig.~\ref{FspecC60}. 
These values are from DFT calculations on 24-atoms
clusters since the same values will be used in our transport calculations. 
The position of these levels suggest that charge transfer between the Au-electrode and molecule
will be relatively week, in comparison with Cu-electrodes where the
molecule can pick up a pronounced negative excess charge.
 
\subsection{Local Density of States}
\begin{figure*}[tbp]
\begin{center}
\begin{tabular}{ccc}
\includegraphics[width=0.3\linewidth]{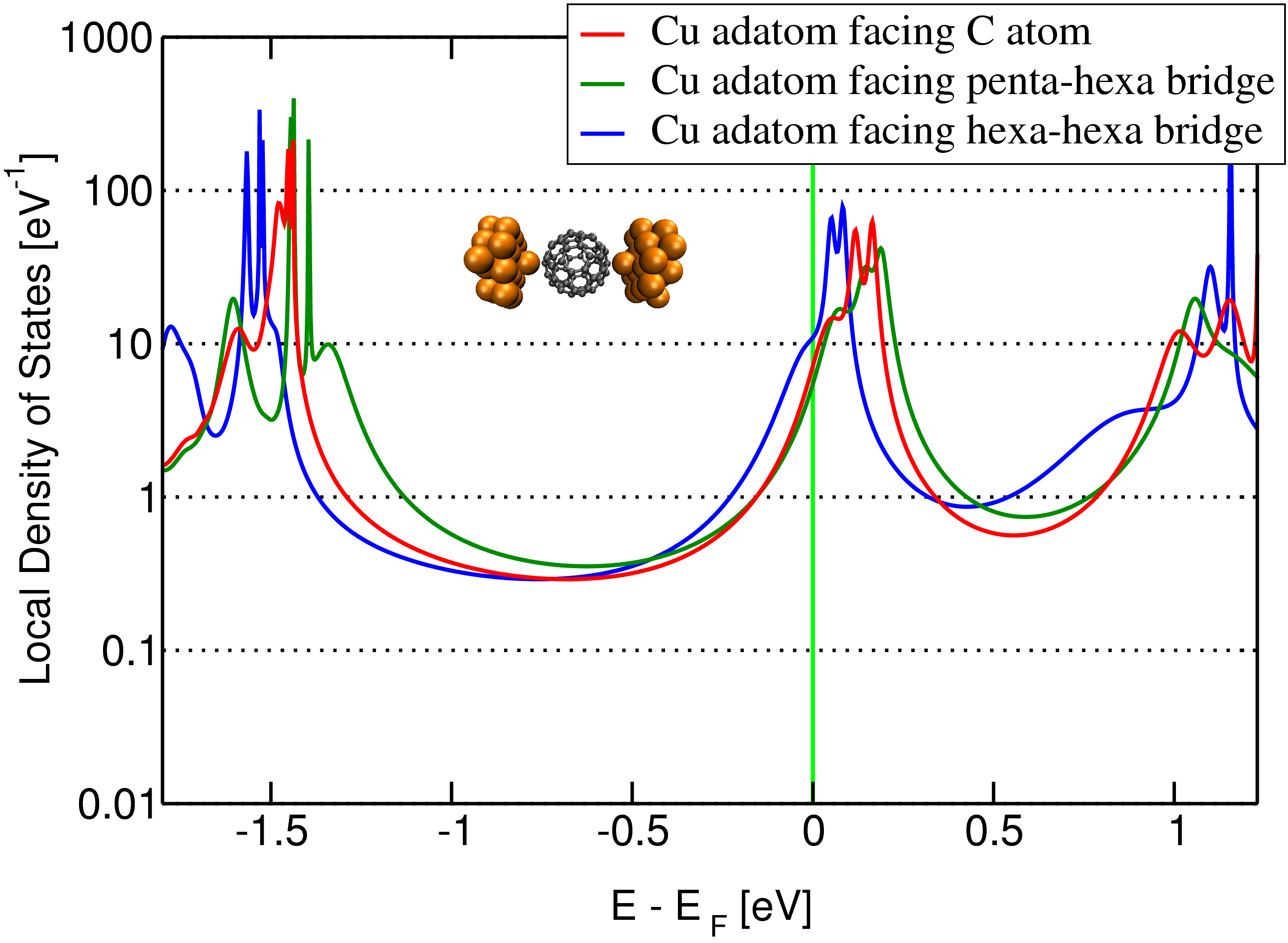} &
\includegraphics[width=0.3\linewidth]{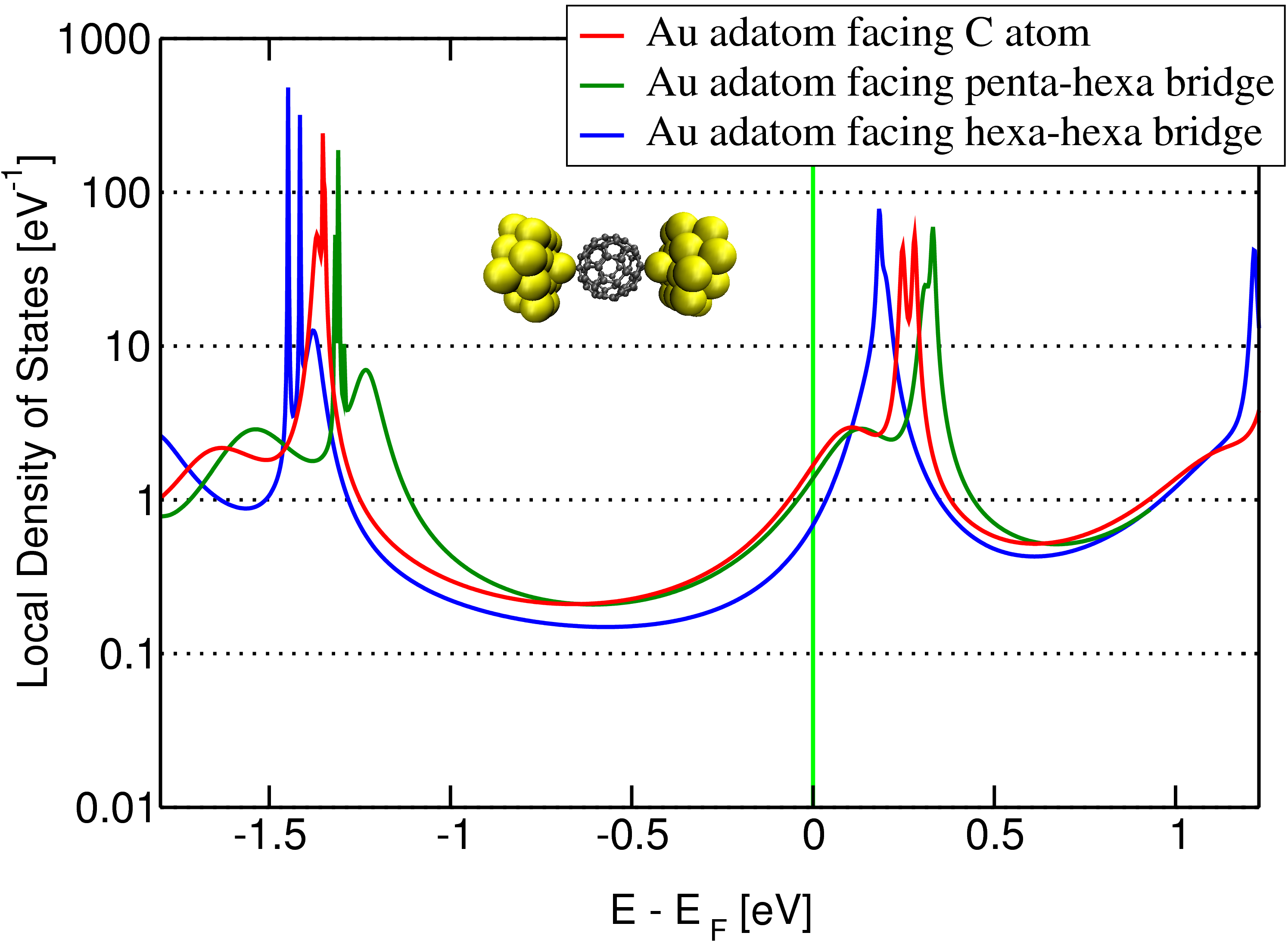} &
\includegraphics[width=0.34\linewidth]{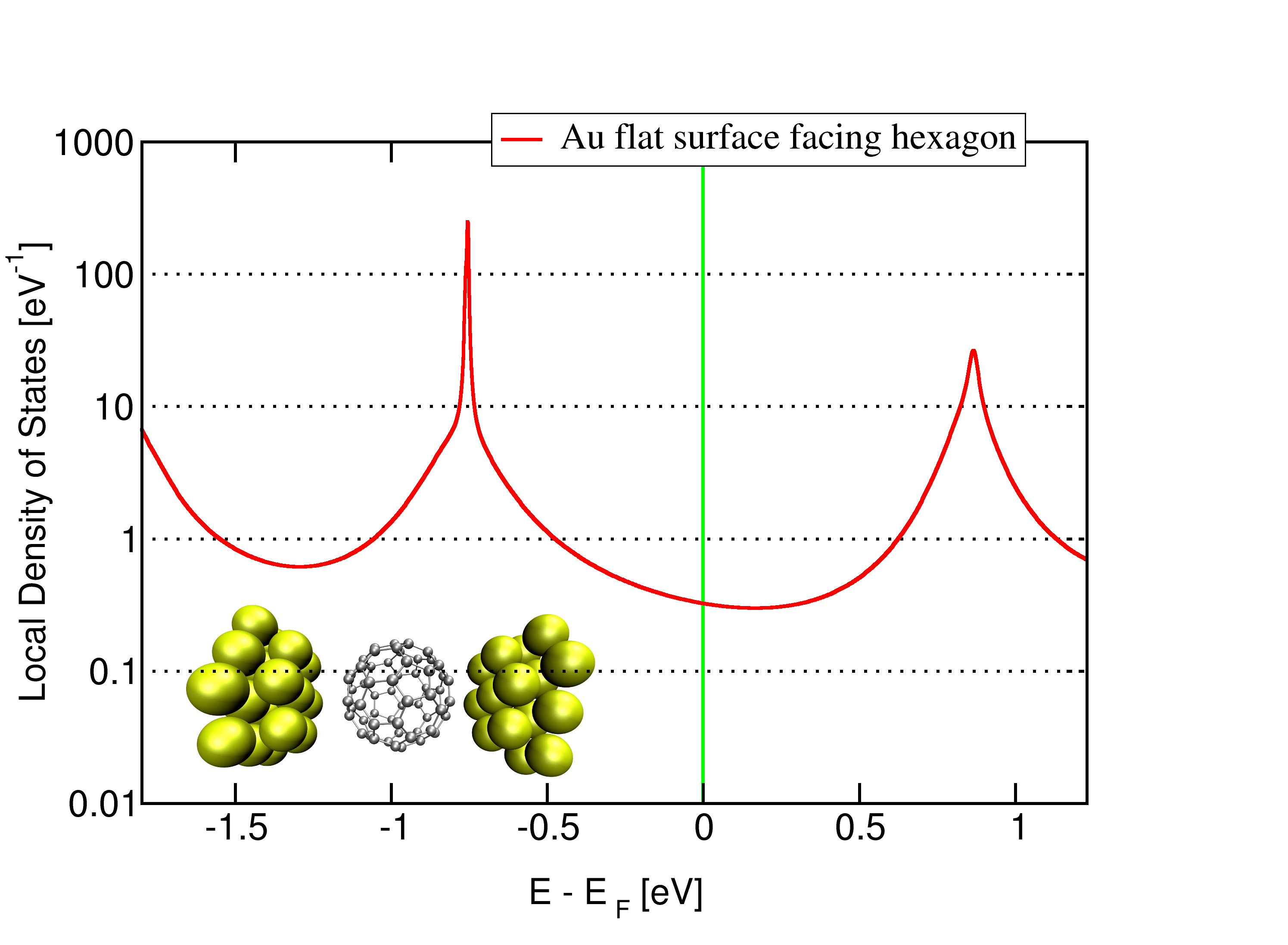} \\
(a) &
(b) &
(c)
\end{tabular}
\end{center}
\caption{(Color) Local Density of States (LDoS) projected on C$_{60}$ for
  Cu-electrodes (a) and Au-electrodes (b) for the three geometries of
  lowest total energies identified in Fig. \ref{f1}.
  (c) LDoS of C$_{60}$ in a junction with direct coupling to a
  flat Au(111) surface for the optimum geometry, see Fig.~\ref{f1}c.}
\label{f3}
\end{figure*}

%% LDOS 
 When the molecule is in contact with metal electrodes, the position of these molecular levels will shift
and experience a lifetime broadening. The 
electronic structure of the junction is represented by the density of
states projected on the C$_{60}$ (local density of states, LDOS).
We calculate it for the geometries of lowest total energy with 
the DFT-based Green's function formalism for non-interacting
particles described in Appendix~\ref{AppB} and Ref. \onlinecite{TranspForm}. 
The result is displayed in Fig. \ref{f3}a,b. 
%%%%%%%%%%%%%%%%%%%%%%%%%%%%%%%%%%%%%%%%%%%%%%%%%%%%%%
%%%%%%%%%%%%%%%%%%%%%%%%%%%%%%%%%%%%%%%%%%%%%%%%%%%%%% 

To quantify further,  we parametrize the LDOS as a sum of
Lorentzians, 
\begin{equation}
\text{LDOS}(E) = \frac{1}{\pi} \sum_{n}^{} \frac{\delta_n}{(E - \epsilon_n)^2 + \delta_n^2}.
\label{ELDOS}
\end{equation}
Values  of the fitting parameters, resonance position $\epsilon_n$ and
broadening $\delta_n$, are given in Table \ref{t1} for junctions where the adatom is 
facing a C-atom of the C$_{60}$ (Fig.~\ref{f1}).
  %%%%%%%%%%%%%%%%%%%%%%%%%%%%%%%%%%%%%%%%%%%%%%%%%%%%%%
\begin{table*}[bhtp]
\begin{center}
\begin{tabular}{|c|c|c|c|c|c|c|c|c|c|c|c|c|}
\hline
 &  & \multicolumn{5}{|c|}{HOMO} & \multicolumn{3}{|c|}{LUMO} & \multicolumn{3}{|c|}{LUMO+1} \\ \hline
\multirow{2}{*}{Au} & $\epsilon_n-E_\text{F}$ (eV) & -1.68 & -1.38 & -1.37 & -1.35 & -1.35 & 0.10 & 0.25 & 0.28 & 1.15 & 1.28 & 1.33 \\\cline{2-13}
 & $\delta_n$ (meV) & 149  & 26 & 8 & 1 & 6 & 116 & 8 & 6 & 175 & 21 & 1 \\ \hline
\multirow{2}{*}{Cu} & $\epsilon_n-E_\text{F}$ (eV) & -1.59 & -1.48 & -1.48 & -1.45 & -1.44 & 0.06 & 0.12 & 0.16 & 1.02 & 1.15 & 1.24 \\\cline{2-13}
 & $\delta_n$ (meV) & 55  & 15 & 17 & 5 & 3.5 & 50 & 12 & 11 & 60 & 40 & 1.1 \\ \hline
\multirow{2}{*}{Au (flat)} & $\epsilon_n-E_\text{F}$ (eV) & -0.78 & -0.77 & -0.76 & -0.75 & -0.75 & 0.84 & 0.86 & 0.87 & 1.92 & 1.92 & 1.93 \\\cline{2-13}
 & $\delta_n$ (meV) & 114  & 116 & 4 & 2 & 2 & 82 & 75 & 17 & 9 & 42 & 53 \\ \hline
\end{tabular}
\end{center}
\caption{Linewidths, $\delta_n$,  and positions of energy levels, 
$\epsilon_n$, for the geometry where the adatom sits on-top of a C-atom, and the flat surface
geometry. Levels are classified  according to their position in the energy spectrum of the
molecule in vacuum (HOMO, LUMO, HOMO+1).}
\label{t1}
\end{table*}
%%%%%%%%%%%%%%%%%%%%%%%%%%%%%%%%%%%%%%%%%%%%%%%%%%%%%%

\subsubsection{Adatom geometry}

The splitting of molecular energy levels in Fig.~\ref{f3}, especially the LUMO ones,  
indicates that the formation of the chemical bond constitutes a significant perturbation
in the sense that one of the LUMO states of the molecule splits away from the others. It hybridizes
more strongly with the electrode states as exhibited by the corresponding increased level
broadening.  Electron transport through 
C$_{60}$ will be mostly via this level. 
Table \ref{t1} reveals the effect of electrode coupling on level
splitting and broadening is about  two times stronger for Au- than for
Cu-electrodes.

%%%%%%%%%%%%%%%%%%%%%%%%%%%%%%%%%%%%%%%%%%%%%%%%%%%%%%
\subsubsection{Flat Au(111) contacts}
%%%%%%%%%%%%%%%%%%%%%%%%%%%%%%%%%%%%%%%%%%%%%%%%%%%%%%

With a flat Au(111) surface (see
Fig.~\ref{f1}c) $E_\text{Fermi}$ resides in the middle of
the HOMO-LUMO gap, much closer to its vacuum position (see Fig.~\ref{FspecC60})
than in the adatom geometries. This indicates, that partial charge
transfer is weaker with a flat electrode as compared to the case of
adatoms, reflecting that binding is purely of the van-der-Waals type.
Consequently, the splitting of the energy levels (see
Table~\ref{t1})
is much smaller than that of the adatom geometry. In other words, 
the presence of the electrodes implies only a weak symmetry breaking
with an overall small effect on the molecular frontiers orbitals. 
We associate the large quantitative differences in the observed shifts
and broadenings  with wavefunction overlaps: 
symmetry related extinction results in reduced hybridization matrix
elements. 

%%%%%%%%%%%%%%%%%%%%%%%%%%%%%%%%%%%%%%%%%%%%%%%%%%%%%%
%%%%%%%%%%%%%%%%%%%%%%%%%%%%%%%%%%%%%%%%%%%%%%%%%%%%%%
\section{Transport calculations for C$_{60}$}
%% transmission

If we were to consider a system with $N$ channels that do not interfere with each 
other, then by definition  the  transmission (per spin)  could be
written as a sum of Lorentzians, 
\begin{equation}
\label{e2}
T(E) = \sum_{n}\ \frac{\delta_n^2}{(E - \epsilon_n)^2 + \delta_n^2}.
\end{equation}
with parameters given in Table \ref{t1}. The formula yields a good
approximation, usually, in the presence of symmetric coupling if 
a single transport resonance dominates. However, in cases where transfer amplitudes of
different  channels can be of comparable magnitude interference terms
may become significant and the approximation (\ref{e2}) breaks down. 
Then, in principle, the Landauer formula, Eq.~(\ref{e18}) in the
appendix, has to be employed. 

%%%%%%%%%%%%%%%%%%%%%%%%%%%%%%%%%%%%%%%%%%%%%%%%%%%%%%
\subsection{Results: Transmission function }
%%%%%%%%%%%%%%%%%%%%%%%%%%%%%%%%%%%%%%%%%%%%%%%%%%%%%%

\begin{figure*}[bht]
\begin{center}
\begin{tabular}{ccc}
\includegraphics[width=0.32\linewidth]{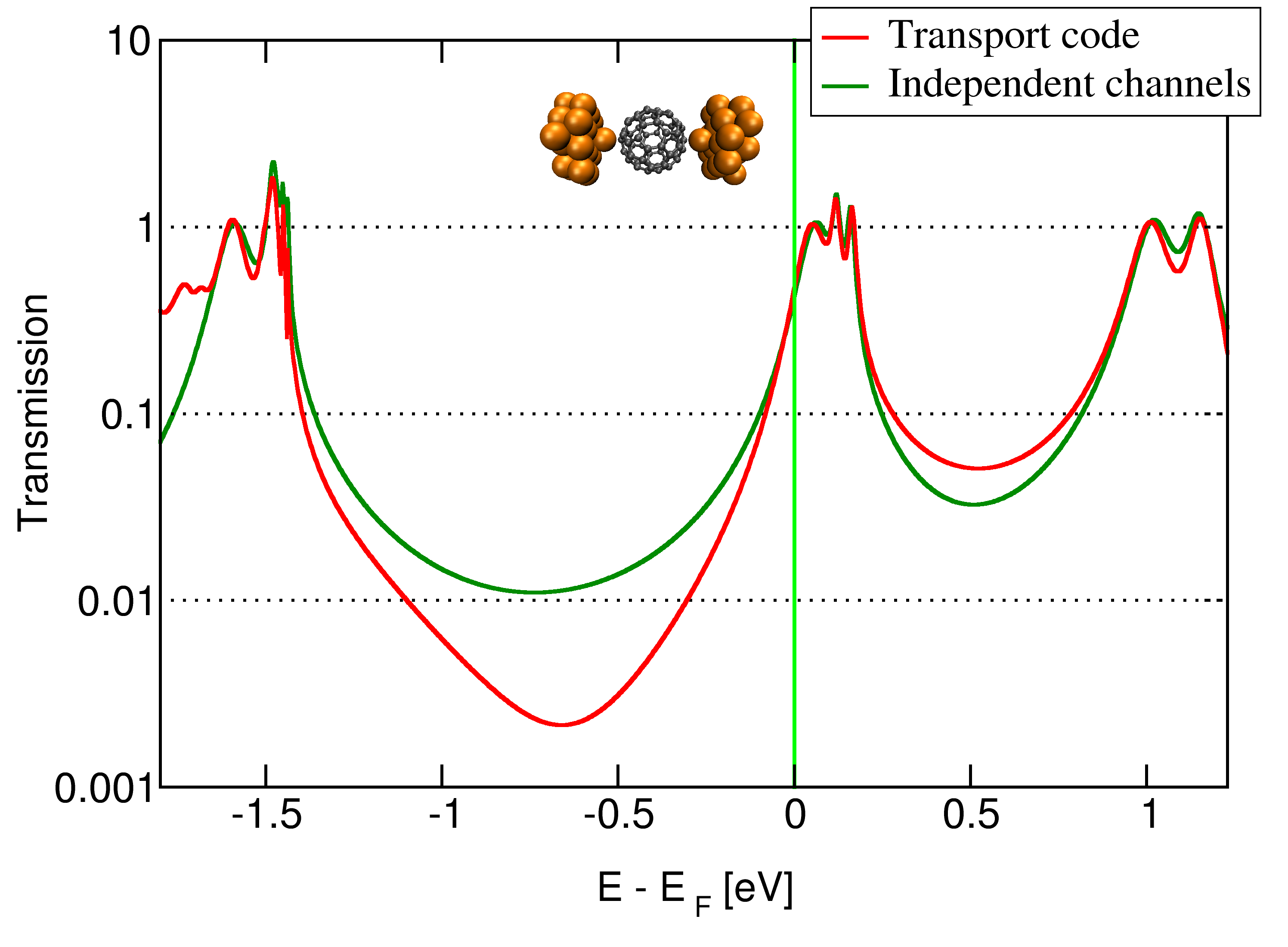} &
\includegraphics[width=0.32\linewidth]{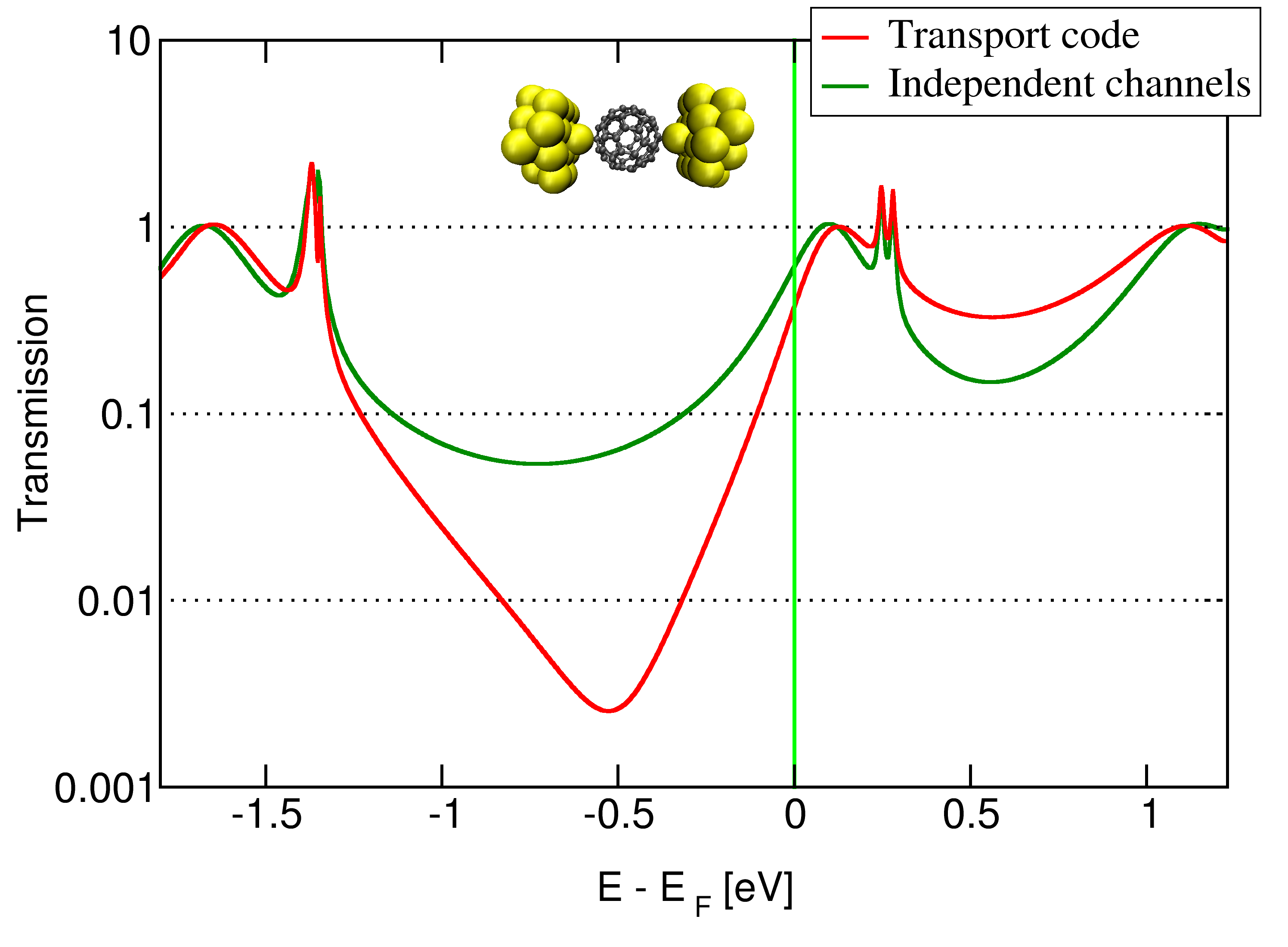} &
\includegraphics[width=0.32\linewidth]{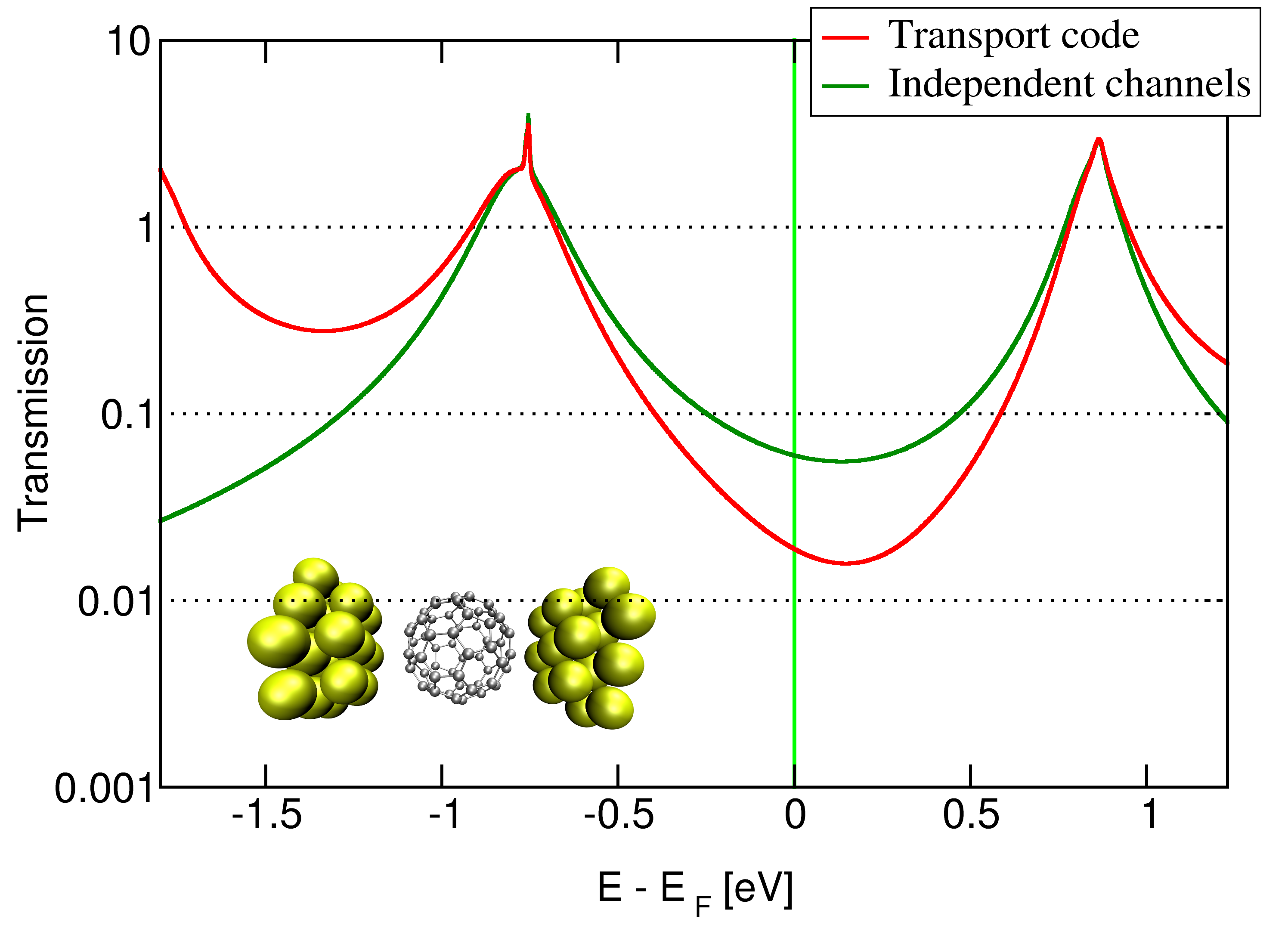} \\
(a) &
(b) &
(c)
\end{tabular}
\end{center}
\caption{(Color) Transmission functions for different electrode models.
  Adatom in atop-position with Cu-electrode (a), and
  Au-electrode (b). (c) diplays the flat Au(111)-surface.
Red line: DFT-based transport code with
Landauer formula/NEGF-formalism. 
Green line: isolated resonances model, Eq.~(\ref{e2}). Plot
highlights the effect of crosstalk between different transport
channels in the transmission valley regime of energies.}
\label{f4}
\end{figure*}

The full transmission function, $T(E)$, as obtained from the Landauer
formula in the NEGF-formulation\cite{TranspForm} 
is displayed in Fig. \ref{f4}a,b  for the Au-C atop-geometry. 
It  exhibits pronounced non-Lorentzian features, with 
a suppression of the transmission in the valley region. 
Comparing with the model of isolated resonances, Eq.~(\ref{e2}), 
one can see that the missing cross-terms between different 
transfer modes explains this behavior. 
We conclude that interference  between transport resonances
play a quantitatively important role
in electron transport through C$_{60}$ when the chemisorbed
C$_{60}$-molecule is only weakly
charged. The conductance could be reduced by roughly one order
of magnitude as a result
% compared to an isolated resonances approximation.

%%%%%%%%%%%%%%%%%%%%%%%%%%%%%%%%%%%%%%%%%%%%%%%%%%%%%%
\subsection{Discussion: effective two level model}
%%%%%%%%%%%%%%%%%%%%%%%%%%%%%%%%%%%%%%%%%%%%%%%%%%%%%%

We interpret our findings for the transmission function using
an effective two level model. Its precise
definition together with a derivation of
basic properties are given in the appendix \ref{AppD}.
\footnote{ Here, we use a slightly simplified version in which we
choose an approximation $\Psi_c=0,\pi$. In this way
we eliminate two fitting parameters, the real angle $\Psi_c$ and
a background contribution that cuts off the Fano-anti-resonance.
Both parameters need to be accounted for if the
simplified (toy) model,
when adjusted near the resonance positions, does not
capture the behavior of the transmission function
in the valley region. }
The salient feature of the simplified model are
summarized by the following set of equations: 
\bea
T(E) &=& T_0+T_1\pm \bar T_{01} \label{e3} \\
\bar T_{01}(E)&\approx& 2\sqrt{T_0T_1}
\frac{ (E-\epsilon_0)(E-\epsilon_1)+\gamma_0\gamma_1}
     {\sqrt{(E{-}\epsilon_0)^2{+}\gamma_0^2}\sqrt{(E{-}\epsilon_1)^2{+}\gamma_1^2}}
\label{e4}
\eea
The two levels that one should refer to here derive from 
the C$_{60}$-HOMO quintett and LUMO-triplett, that exhibit
resonances with the strongest broadening. Accordingly, we see from
Table \ref{t1} the set of parameters: $\epsilon_0=-1.68$eV,
$\epsilon_1=0.1$eV, $\gamma_{0}=0.149$meV, $\gamma_1=0.116$meV.
A decomposition of the LDoS given in Appendix \ref{AppC} further
substatiates this simplification. 
%%%%%%%%%%%%%%%%%%%%%%%%%%%%%%%%%%%%%%%%%%%%%%%%%%%%%%%%%%%%%%%%%%%%%%%%%%%
\begin{figure*}[tbp]
\begin{center}
\begin{tabular}{ccc}
\includegraphics[width=0.32\linewidth]{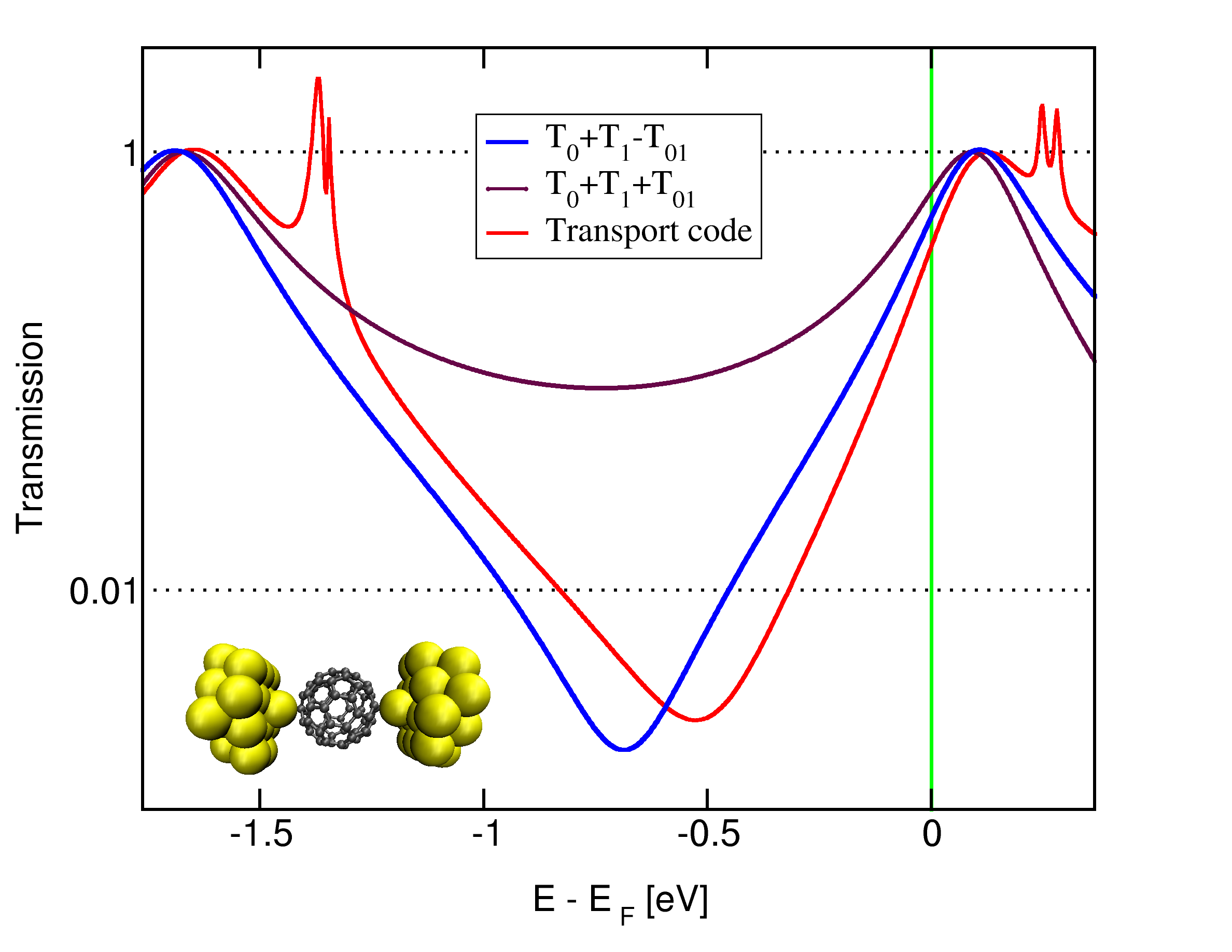} &
\includegraphics[width=0.32\linewidth]{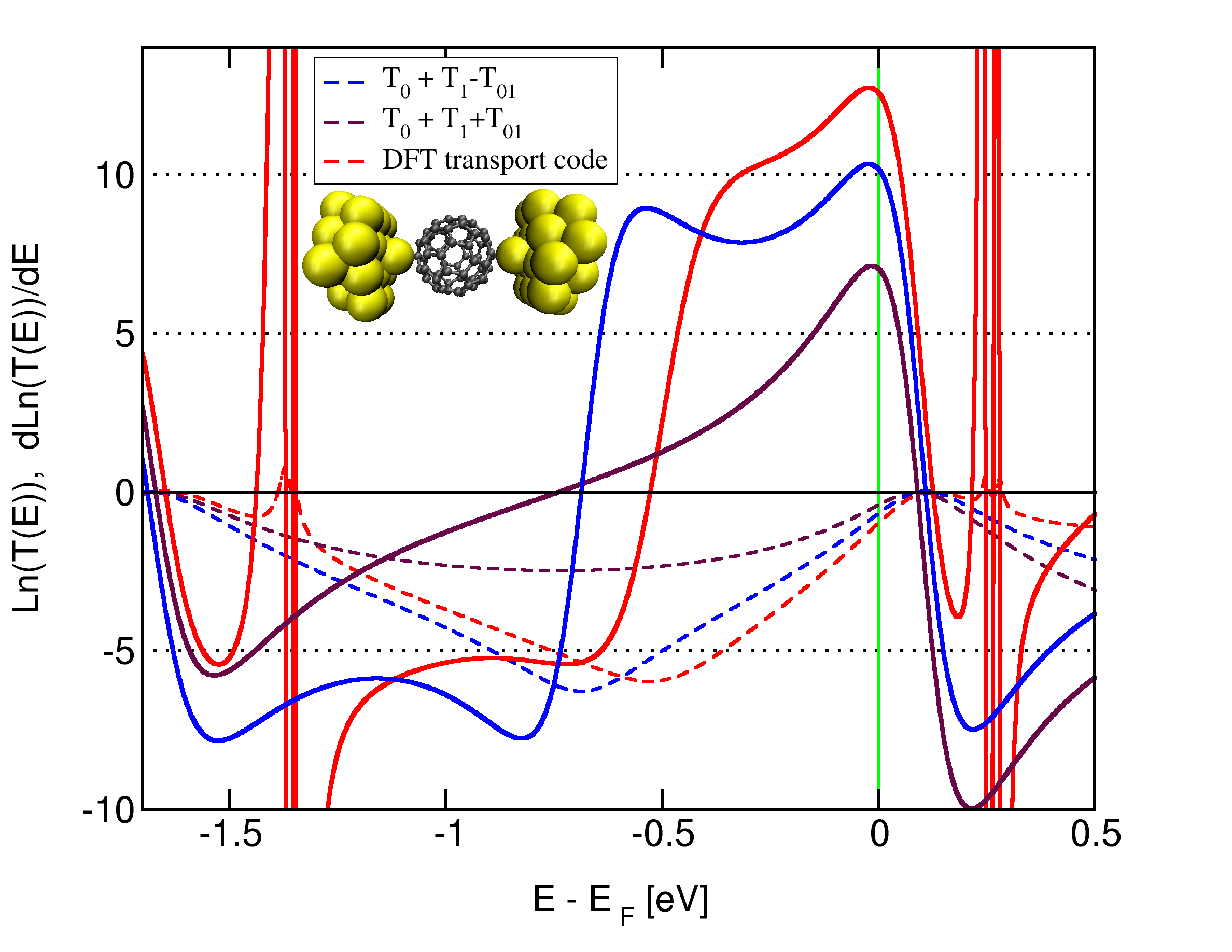} & 
\includegraphics[width=0.32\linewidth]{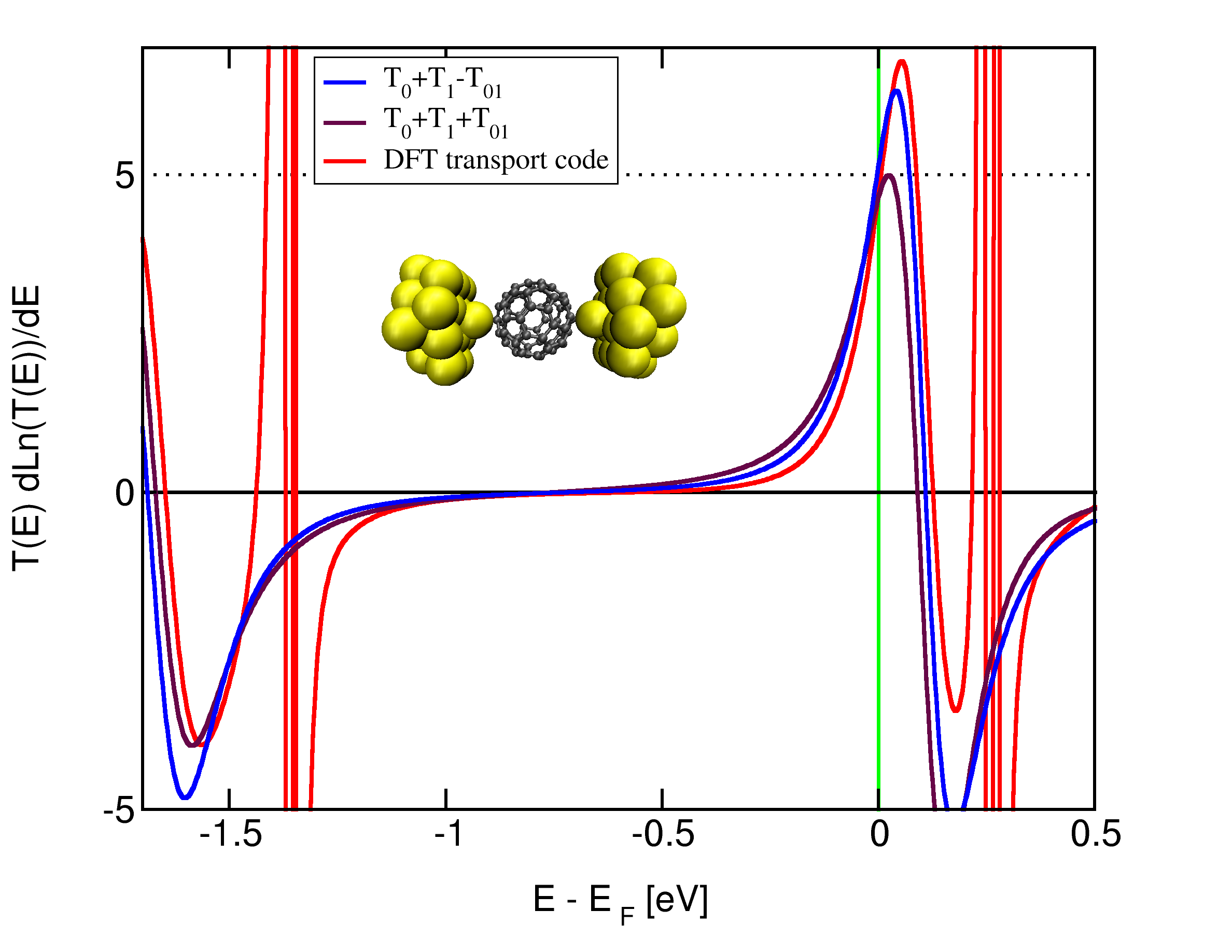} \\
(a) &
(b) &
(c)
\end{tabular}
\end{center}
\caption{(Color) (a) Comparison of transmissions obtained from the
  transport code (Fig. \ref{f4}(b), solid red)
  and the two-states model
  (solid blue), see Eq.~(\ref{e3}).
  In order to highlight the impact of the interference term, 
$\bar T_{01}$, Eq.~(\ref{e3}),  results for constructive and destructive situations are
given.
  (b) Logarithm of the transmission function (dashed lines) and its
  derivative. The latter represents the system specific
  information content of the Seebeck-coefficient, Eq.~(\ref{e7}). 
  (c) System specific, energy dependent characteristics, $T(E)d\ln
  T(E)/dE$, of the thermal current Eq.~(\ref{e6}). Plot highlights
  a result of the two-level model: traces for destructive (blue)
  and constructive (brown) interference
  give nearly coinciding results, even though their electronic
  transmission deviates by orders of magnitude, see the transmission
  functions given in the left figure.   
 }
\label{f5}
\end{figure*}
%%%%%%%%%%%%%%%%%%%%%%%%%%%%%%%%%%%%%%%%%%%%%%%%%%%%%%%%%%%%%%%%%%%%%%%%%%%%

The sign in Eq.~(\ref{e3}) controls
the effective mixing between
the two transport channels, i.e. whether they interfere constructively
(minus sign) or destructively (plus sign) energies in the valley
region, $\epsilon_0<E<\epsilon_1$. As we explain in the appendix, destructive
interference occurs in two-level models where both states couple with similar strength to both reservoirs.
For C$_{60}$, we expect that there should not be any important 
difference between the coupling of the HOMO and LUMO 
levels to the leads, we can expect to see destructive interference. 

%Since for  C$_{60}$ there is no reason to assume that with
%respect to the lead-coupling there should be a qualitative difference
%between the salient HOMO- and LUMO-levels, one expects that
%C$_{60}$ belongs to this class.
Indeed, already from Fig. \ref{f4} we can see that for the case of
the C$_{60}$-Au-junction, the transmission in the valley region is very
strongly suppressed supporting our claim. 

Hence, we conclude that the plus-sign
should be chosen in Eq.~(\ref{e3}). Furthermore, with symmetric
coupling we also have 
\be
\label{e5}
T_i(E) = \frac{\gamma_i^2}{(E-\epsilon_i)^2+\gamma_i^2}, \qquad i=0,1.\\
\ee
The sign in Eq.~(\ref{e3}) is the toy model's only
ingredient in $T(E)$ that is not fixed by the LDOS alone. 

In Fig.~\ref{f5} (left) we compare the
full transmission with the one
from the toy model, Eq.~(\ref{e3}).
Indeed, in the valley region
the toy model reproduces the transmission and all its 
non-Lorentzian features well up to a 
 small shift of the minimum-transmission energy.
This shift is readily explained, e.g., by residual energy dependencies in the
pole-positions due to the structured density of states in the
reservoirs. For the minimum conductance at energy $\epsilon^*$
we obtain a parametrical estimate $T^*_\text{constr}\approx
4 \gamma_0\gamma_1/|\epsilon_0-\epsilon^*||\epsilon_1-\epsilon^*|$
for constructive interference and 
\be
T^*_\text{destr}\approx\frac{1}{4} {T^*}^2_\text{constr}
\ee
in the other case.

%%%%%%%%%%%%%%%%%%%%%%%%%%%%%%%%%%%%%%%%%%%%%%%%%%%%%%%%%%%%
\subsection{Thermopower}
%%%%%%%%%%%%%%%%%%%%%%%%%%%%%%%%%%%%%%%%%%%%%%%%%%%%%%%%%%%%
We complete our account of C$_{60}$
transport properties with a discussion of the thermopower of an Au-C$_{60}$-Au junction.
The thermopower or Seebeck coefficient $S$ determines the magnitude of
the built-in potential developed across the junction
when a temperature difference $\Delta \mathfrak{T}$ is applied.
\cite{imry86}
In particular, molecular junction thermopower can be useful in
determining the dominant
molecular orbital for transport and the
identity of the primary
charge carriers.
\cite{ke09,widawsky12}
With the additional presence of an external voltage bias $\Delta V$ across
the junction\cite{paulsson03,segal05}, the total current $I$ in this case is simply
\be
\label{e6}
I = \frac{2e^2}{h} \left[-T(E_\text{F})\Delta V + T(E_\text{F}) S(E_\text{F}) \Delta \ \mathfrak{T}\right]
\ee
%\be
%\label{e6}
%I = S(E_\text{F})\  T(E_\text{F}) \Delta \mathfrak{T}\ \left[\frac{2e^2}{h}\right], 
%\ee
where $T(E_\text{F})$ denotes the transmission at the Fermi energy. 
For non-interacting electrons, the Seebeck-coefficient, $S(E)$,
is closely related to the transmission function $T(E)$,
ultimately, because all transport processes are controlled by
the tunneling probabilities of electrons with a given energy
through the barrier. We have (using the convention $e = |e|$)
\be
\label{e7}
S(E) = \frac{\pi^2 k^2_\text{B}\mathfrak{T}}{3e} \frac{\mathrm{d} \ln T(E)}{\mathrm{d}E}
\ee
where $\mathfrak{T}$ denotes the (electronic) temperature. The system
specific information is all encoded in the logarthmic derivative,
which we now discuss. 

Fig. \ref{f5}b displays the logarithmic derivative for our model
system, C$_{60}$. Again, we can convince ourselves that the
two-level model with destructive interference
accounts well for the salient features of the
full DFT-based trace. A striking characteristics to be observed here
is the step-like transition, changing sign, that the derivative undergoes when the
energy sweeps by the valley minimum point $\epsilon^*\approx
-0.51$eV. We can estimate the parametrical dependency of the
slope at $\epsilon^*$ employing Eqs.~(\ref{e3}-\ref{e5}):
\be
\left. \frac{d\ln T(E)}{dE}\right|_{E\approx \epsilon^*}
\approx
\frac{(E{-}\epsilon^*)(\epsilon_1{-}\epsilon_0)^2}{2\gamma_0\gamma_1|\Delta_0\Delta_1|},
\quad |E-\epsilon^*|\apprle \gamma_{0},\gamma_1
\ee
where $\Delta_{i}=\epsilon_i-\epsilon^*$. For the typical cases
where $\gamma_0$ and $\gamma_1$ are comparable, $\epsilon^*$ is
halfway between $\epsilon_1$ and $\epsilon_0$, so the expression
simplifies further:
$d\ln T(E)/dE|_{\epsilon^*} \approx
2(E-\epsilon^*)/\gamma_0\gamma_1$.
Hence, the slope diverges in the weak coupling limit,
where $\gamma_{0,1}$ tend to zero at fixed level splitting
$|\epsilon_{1}-\epsilon_{0}|$.

The estimate for the slope is valid for energies in a vicinity
of width $\gamma_{0,1}$ about $\epsilon^*$. Hence, the logarithmic
derivative takes very large magnitudes $\sim \pm 1/\gamma_{0,1}$
near the center of the valley region. It is only near the resonances
where it reaches similar values, e.g. $\sim 1/\gamma_{1}$ near
$\epsilon_{1}$. The intermediate region interpolates between
these to maxima.

This behavior is typical of systems that exhibit almost perfect
destructive interference, so that $T(E)$ approaches zero near
some energy $\epsilon^*$. It is completely absent with constructive
interference, see Fig. \ref{f5} (center).
There the logarithmic derivative has a parametrically small slope of the
order of $\sim 1/|\Delta_0\Delta_1|$ (rather than $1/\gamma_0\gamma_1$)
near $\epsilon^*$. 

The temperature driven current is, up to system unspecific
prefactors, given by the product $T d\ln(T)/dE$.
Each factor has been seen to be very sensitive to
the sign in Eq.~(\ref{e3}) in our discussion.
For the product this is not the case, as one infers from
Fig. \ref{f5}c. The reason is that the suppression of
$T^*$ for the case of destructive interference is largely 
compensated by the strong slope in the logartithmic derivative.
Indeed, in the valley region we have for the product with
either destructive of constructive interference a similar
behavior: $T^* d\ln(T)/dE \approx
(\epsilon-\epsilon^*)\gamma^2/\delta^4$,
where $\gamma=\gamma_0=\gamma_1,
\delta=\Delta_0=-\Delta_1=(\epsilon_0-\epsilon_1)/2$
has been assumed. 

Hence, we arrive at the following conclusion:
In the presence of destructive interference (minus sign in
Eq.~(\ref{e3})) the bias voltage driven
current can be suppressed by orders of magnitude in the valley
region. Nevertheless, an electrical current driven by a thermal
bias reaches similar values as it would in the
absence of interference effects.

%%%%%%%%%%%%%%%%%%%%%%%%%%%%%%%%%%%%%%%%%%%%%%%%%%%%%%%%%%%%%%%%%%%%%%%%%%%
%%\begin{figure}[tp]
%%\begin{center}
%%\includegraphics[width=1\columnwidth]{ModelTransmission_TdLnT}
%%\end{center}
%%\caption{System specific, energy dependent characteristics, $T(E)d\ln
%  T(E)/dE$, of the thermal current Eq.~(\ref{e6}). The two-level
%  traces for destructive (blue) and constructive (brown) interference
%  give nearly coinciding results, even though their electronic
%  transmission deviates by orders of magnitude, see Fig.~(\ref{f5}). 
% }
%\label{fLog}
%\end{figure}
%%%%%%%%%%%%%%%%%%%%%%%%%%%%%%%%%%%%%%%%%%%%%%%%%%%%%%%%%%%%%%%%%%%%%%%%%%%

%%%%%%%%%%%%%%%%%%%%%%%%%%%%%%%%%%%%%%%%%%%%%%%%%%%%%%%%%%%%
\subsection{Contact geometries revisited: asymmetry and chain formation}
%%%%%%%%%%%%%%%%%%%%%%%%%%%%%%%%%%%%%%%%%%%%%%%%%%%%%%

In this section we show, that
the shape of the transmission function can be significantly modified
by changing the contact geometry. 

\subsubsection{Single adatom contacts breaking inversion symmetry}

\begin{figure}
\begin{center}
\includegraphics[width=1\columnwidth]{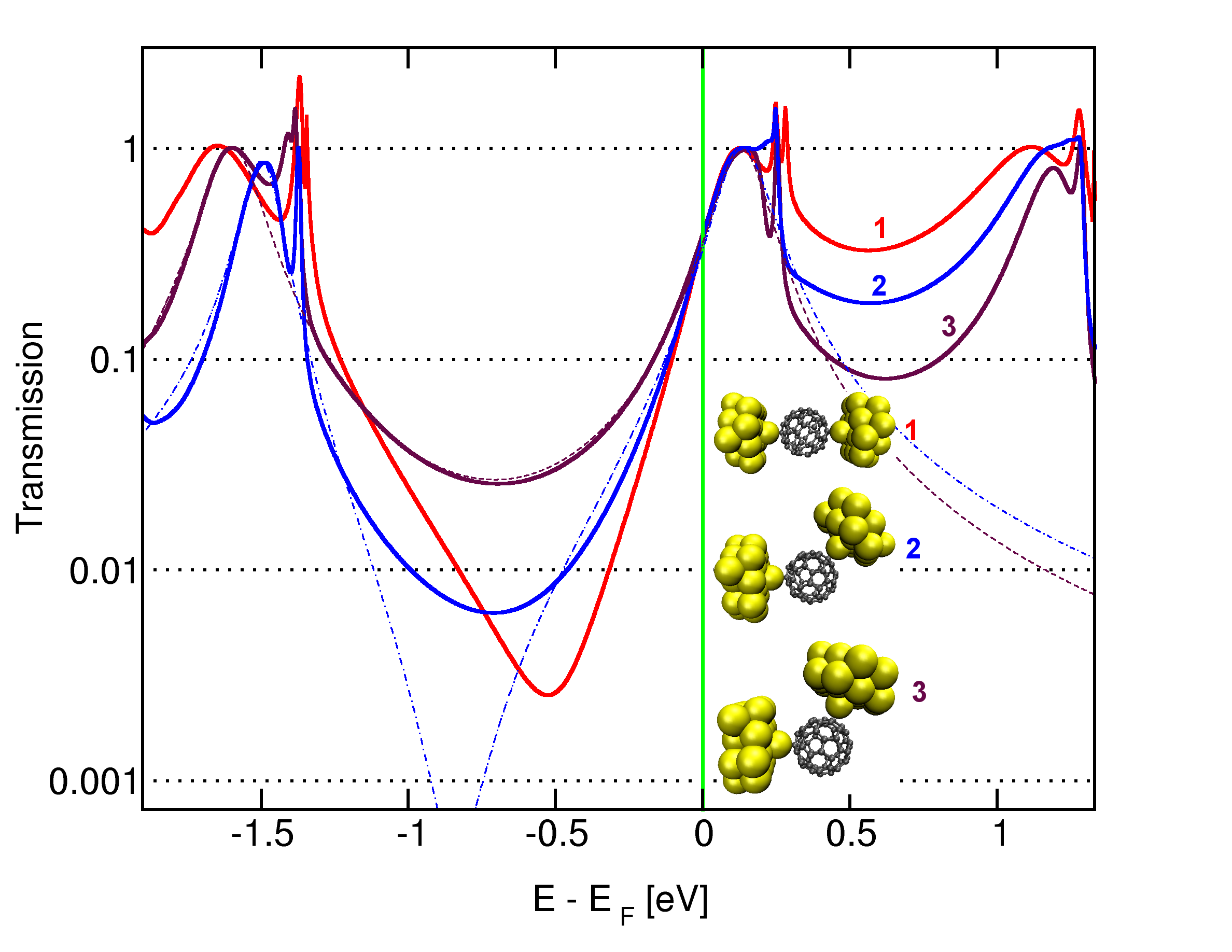}
\end{center}
\caption{Transmission function for different
  anchoring geometries: (1) Symmetrical geometry
  from previous plot Fig. \ref{f5}.
  Asymmetrical geometries (2, maroon; 3, blue).
  Position (2) can be fitted by
 adding two Lorenzians (dashed line, $\gamma_0=0.106, \epsilon_0=-1.6,
 \gamma_1=0.096,\epsilon_1=0.13$), indicating that interference effects are
 weak. In contrast, geometry (3) exhibits destructive interference.
 The model (\ref{e3}) fails in the valley region (blue dot dashed line),
 mainly because the angle $\Psi_c$ (defined in Appendix \ref{AppD})
 is not close to $\pi$. From the blue data trace one
 estimates roughly $\cos\Psi_c \approx -0.8$ together with an asymmetry
 $\gamma_{0\text{L}}/\gamma_{0\text{R}}\approx 0.45$.}
\label{f6}
\end{figure}

By selecting different pairs of C-atoms on the molecule,
one modifies the phase difference between parallel transmission paths
(Fig. \ref{f6}). The situation is similar to the case considered 
perviously\cite{ke08,cardamone06}, except that there a torus geometry 
was considered while we investigate a sphere. 
%LV: Cite Baranger, Stafford, ....
The different contact geometry has an impact on the effect of $T_{01}$.
Specifically, the transmission function for the second geometry
(maroon trace in Fig. \ref{f6}) is very similar
to a pure superposition of Lorentzian resonances, suggesting that
the $T_{01}$ in Eq. \ref{e3} does not contribute significantly and
interference effects between parallel paths are much weaker
than in the other cases.

In contrast, destructive interference reappears in geometry (3),
Fig. \ref{f6}. This geometry is however, far from perfectly symmetric,
and therefore the simplified model Eq.~(\ref{e4}) cannot be expected
to hold. Instead, in principle the more complete formula given in the appendix,
(\ref{eD13}-\ref{eD15}), should be applied. Indeed, the fit based on
the simplified expression (\ref{e3}) does not properly reproduce
the behavior of $T(E)$ in the valley region. 
(Based on the discrepancy one expects $\cos(\Psi_c)\approx -0.8$.)

We emphasize that there is a very
large variability of the conductance in the valley region
even though all electrode positions (1-3) are associated with
similar resonance positions and broadenings. 
We attribute this behaviour to the mixing angle $\Psi_c$ oscillating from
$\pi$ (geometry 1) to $0$ (2) back to larger values $\Psi_c\approx
2.5$ (3) again. This observation we take as support for
our claim that
the variations observed in the transmission function are due to
a modification of the phase difference between parallel paths.

%%%%%%%%%%%%%%%%%%%%%%%%%%%%%%%%%%%%%%%%%%%%%%%%%%%%%%%%%%%%%%%%
\subsubsection{Au-contact chains}
%%%%%%%%%%%%%%%%%%%%%%%%%%%%%%%%%%%%%%%%%%%%%%%%%%%%%%%%%%%%%%%%

In order to investigate the development of interference
with decreasing level broadening, we consider here geometries where the molecule
is included between Au-chains. In this configuration, the number of incoming and
outgoing lead channels is limited to essentially a single one.
This reflects in the local density of states at the chain terminating
Au-atom that the molecule couples to. It is more strongly structured
as compared to the case with a single Au-adatom, only; the number of
states that are ready to hybridize with the C$_{60}$-orbitals is
reduced. As a consequence, broadening of molecular orbitals contacting 
Au-chains is in general weaker, and also more complicated since a
convolution of two structured functions (LDoS on molecule and
contact-atom/Au-wire) is involved. 

\begin{figure}
\begin{center}
\includegraphics[width=1\columnwidth]{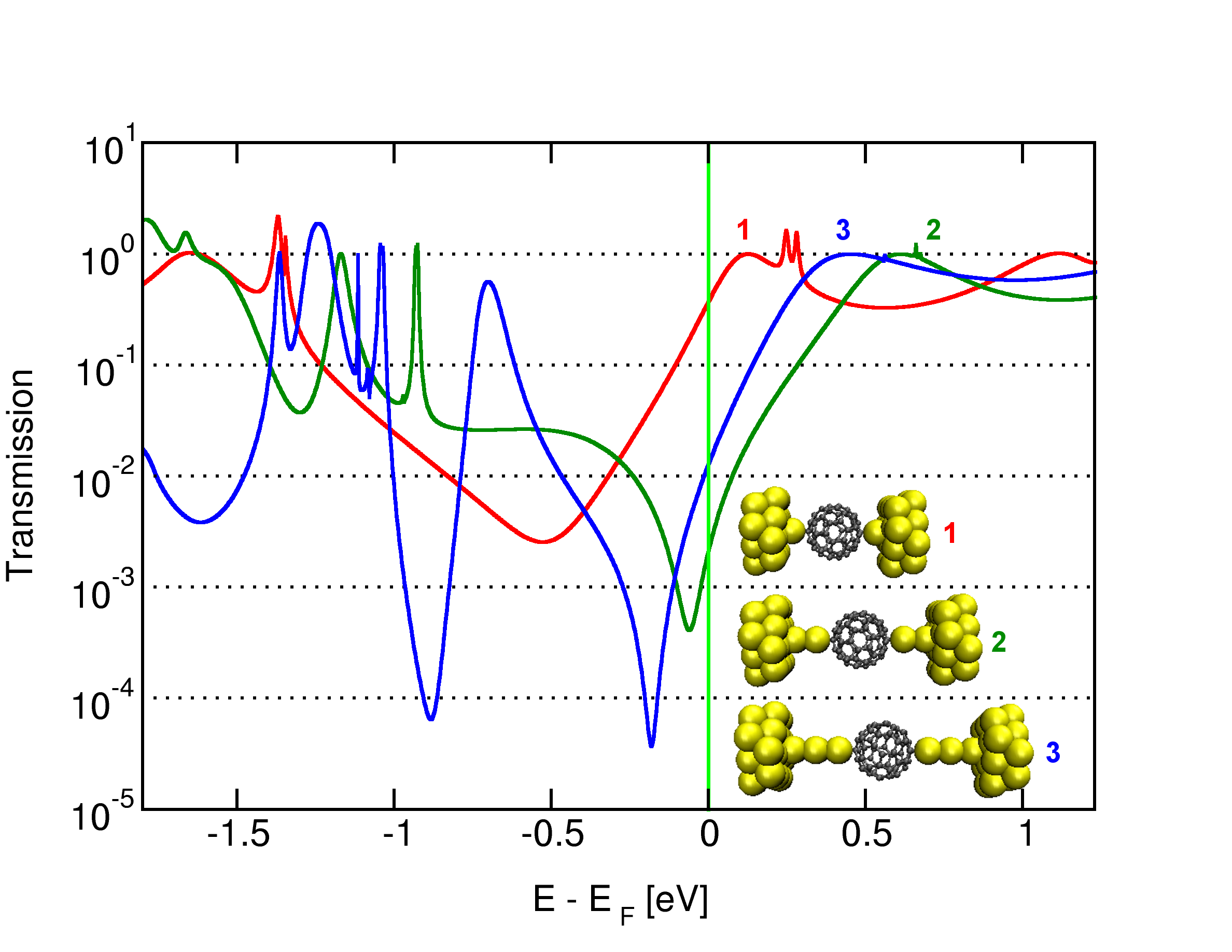}
\end{center}
\caption{Transmission functions for contacts made via adatoms (1),
  two-atoms Au-chains (2) and three-atoms Au-chains (3).}
\label{FTEChains}
\end{figure}

The transmission functions obtained for one, two and three-atoms
chains geometries shown in Fig.~\ref{FTEChains} support these
expectations. We observe a progressive development of large
amplitude anti-resonances with increasing chains length.
They reflect the fact molecular states and wire states
can cooperate in a complicated manner which allows, in particular,
for more levels to develop interference patterns in valley regions. 
In this way, transmission values below $10^{-4}$ can come 
about for 3-atoms chains, which suggests that extremely low
conductance values, between $10^{-3} G_0$ and $10^{-4} G_0$, 
see Fig. \ref{FTEChains}, can
be observed with this type of junctions. This finding becomes
particularly interesting in view of the fact, that a single Au-chain
is well known to exhibit a single perfectly transmitting channel.
Our result Fig. \ref{FTEChains} gives a solid demonstration,
that due to quantum effects even a perfect conductor
can be a very invasive means to fascilitate an electrode coupling.

\subsection{\label{ssIVE} Additional remarks}
We add several remarks on artifacts of DFT-based
transport calculations and about experiments.

First, the functionals used in our study are
well known to underestimate the true HOMO-LUMO gap, as was pointed
out already in section \ref{sIIIA}. The missing derivative
discontinuity influences the alignment of the molecule-based and
metal-based electronic levels; in general it leads to an
overestimation of charge transfer. Hence, while one expects that the
qualitative features of the DFT-based transport calculations are
captured correctly, the positioning of the LUMO with respect to the
Fermi-energy, $E_\text{F}$,  should be slightly too close. The true
$E_\text{F}$ is probably situated somewhat closer to the valley region
than seen in the DFT-calculation. 

Second, recent research has shown that DFT-based transport
calculations employing exact functionals reproduce the exact
transmission for interacting {\it single} level models.
\cite{bergfield11,stefanucci11,schmitteckert11} At present, a rigorous
generalization of the statement to models
with several levels does not exist. In constrast,
the numerical results of Ref. \cite{schmitteckert08}
show deviations between exact conductances
and DFT-based transport with exact functionals in the valley region
indicating that a precise generalization to two-level models
may in fact not exist.

In view of this problem, it is important to realize that the
main finding of our paper is likely to be insensitive
to (weak) interaction effects beyond our GGA treatment.
The reason is, that the statements already follow from a
two-level model with basic ingredients
closed shell system (i.e. no magnetism), 
time-reversal symmetry, inversion symmetry and the fact,
that the HOMO- and LUMO- orbitals are conjugated,
coupling well in similar ways to both leads. 

%For this reason we believe that a direct comparison of
%our analysis with the recent experiment by Yee {\it et al.}
%\cite{yee11} should be meaningful. Due to the slight uncertainty
%in the alignment of LUMO and Fermi level,
%it is no surprise that
%the measured conductance, e.g. $G\sim 3\times 10^{-4}G_0$
%in Ref. \onlinecite{yee11}, somewhat undershoots
%our DFT-estimate, e.g. see Fig. \ref{f6}.
Due to the large, interference induced slope of $T(E)$
in the valley region, one could expect to observe very large
conductance fluctuations in the experiments
-- despite of the high molecular symmetry of C$_{60}$ -- 
due to weak environmental capacitive couplings.
Our study would also suggest that the Seebeck coefficient
does not exhibit such strong fluctuations because
its dependency on the level alignment in the valley region
is relatively weak, see e.g. Fig. \ref{f5} center.
Both qualitative features are indeed observed in
Ref. \onlinecite{yee11}. On a quantitative level, we
observe that the theoretical estimate for the Seebeck-coefficient
overshoots the experimental one roughly by a factor of 2-4.

%%%%%%%%%%%%%%%%%%%%%%%%%%%%%%%%%%%%%%%%%%%%%%%%%%%%%%
%% 
%%%%%%%%%%%%%%%%%%%%%%%%%%%%%%%%%%%%%%%%%%%%%%%%%%%%%%
\section{Conclusions} 

We have presented a detailed study of charge transport
properties of the C$_{60}$-molecule coupled to Au-electrodes.
Our main finding is that the electrical conductance of the molecule
is strongly suppressed due to two interfering transmission channels.
The phenomenon was interpreted as a precursor to a
Fano-anti-resonance. This result has been established by combining
ab-initio transport calculations with a toy-model analysis.
This analysis also suggests, that a thermally driven current is
significantly less sensitive to such interference effects
due to cancellation effects in the transmission function and the
Seebeck coefficient.

\section*{Acknowledgements}
Support by the CFN and SPP 1243 are gratefully acknowledged. Also, 
LV thanks the Packard Foundation for support.

%%%%%%%%%%%%%%%%%%%%%%%%%%%%%%%%%%%%%%%%%%%%%%%%%%%%%%%%%%%%%%%%%%%%%%%%%%%%%%%%%

\appendix
\section{Coefficients for the Grimme correction}
\label{AppA}

We employ the Grimme empirical correction (Eq. \ref{EEdisp}, \ref{Efdmp} and \ref{ECij}) \cite{Grimme2006} to the total GGA energy in order to take 
van-der-Waals interactions into account. The coefficients that we used are given in Table \ref{TableGrimme}. For C and Cu-atoms, the values for 
$C_6$ and $R_0$ have been taken from
Ref. \onlinecite{Grimme2006}. For gold atoms, the $R_0$ coefficient has been obtained from the radius of the 
electron density contour of a single gold atom
%,calculated using a mean-field approach
and the $C_6$ coefficient has been obtained from
a fit to data obtained from second order Moller-Plesset perturbation
theory (MP2) \cite{MatthiasPeter}.
%The scaling factor of 1.1 that was proposed
%by \cite{Grimme2006} has not been used for the $R_0$ coefficient of
%gold because it is not clear whether this scaling factor is relevant for gold.
Following Grimme\cite{Grimme2006} we used $d = 20$ and $s_6 = 1.05$.\\

\begin{equation}
E_{\text{disp}} = -s_6 \sum_{i=1}^{N_{at}-1} \sum_{j=i+1}^{N_{at}} \frac{C_6^{ij}}{R_{ij}^6}f_{\text{dmp}}(R_{ij})
\label{EEdisp}
\end{equation}

\begin{equation}
f_{\text{dmp}}(R_{ij}) = \frac{1}{1 + e^{-d(R_{ij}/R_0 - 1)}}
\label{Efdmp}
\end{equation}

\begin{equation}
C_6^{ij} = \sqrt{C_6^{i}C_6^{j}}
\label{ECij}
\end{equation}

\begin{table}[htbp]
\begin{center}
\begin{tabular}{|c|c|c|}
\hline
   & $C_6$ [J.nm$^6$.mol$^{-1}]$ & $R_0 [\text{\AA}]$ \\ \hline
C  & 1.75	& 1.32 \\ \hline
Cu & 10.8	& 1.42 \\ \hline
Au & 21.9 	& 1.58 \\ \hline
\end{tabular}
\end{center}
\caption{Coefficients used for Grimme empirical correction.}
\label{TableGrimme}
\end{table}

\section{Transport code}
\label{AppB}

A detailed description of our transport simulations has been given in
Ref. \onlinecite{TranspForm}. We present a brief summary of the main steps. 
The effective Kohn-Sham (KS) Hamiltonian $H_{\text{KS}}$ of the extended molecule is  
constructed from the KS-orbitals and energies previously 
computed by DFT (employing the TURBOMOLE package\cite{turbomole89} in our case). 
The Green's function of the extended molecule is built from this Hamiltonian $H_{\text{KS}}$ 
\begin{equation}
G(E) = \frac{1}{E-H_{\text{KS}}-\Sigma}
\label{EG}
\end{equation}
where $\Sigma$ is the self-energy of the reservoirs. It can be taken
in  the form\cite{TranspForm}
\begin{equation}
\Sigma_{nm} = 
\begin{cases}
\delta_{nm}(\delta\epsilon - \mathfrak{i}\eta) & m,n \in S \\
0 & m,n \notin S.
\end{cases}
\label{ESIGMA}
\end{equation}
This self-energy is diagonal in the atomic basis and has non-zero values only in the subspace $S$ associated to the outermost atomic 
layer of each electrode. The parameter $\eta$ = 2.72 eV is the leakage
parameter. It  is adjusted such a way that the transmission functions
are (approximately) invariant
under variation of $\eta$. 
%{\bf FE: Discuss with AB.}
The energy shift $\delta\epsilon$ is tuned for each calculation so that the Fermi level for the entire system 
remains unchanged compared to what has been obtained from DFT 
calculation, i.e. about the Fermi energy of the metal used for the electrodes. 
Here, we used 
$-1.25 \text{eV} < \delta\epsilon < -1.15
\text{eV}$  for Au and 
$-2.00 \text{eV} < \delta\epsilon < -1.94 \text{eV}$ for Cu.

The local density of states (LDOS) projected on the molecule can then be written
\begin{equation}
\text{LDOS}(E) = -\frac{1}{\pi} \sum_{n \in M} \Im~G_{nn}
\label{Appendix:ELDOS}
\end{equation}
where $M$ is the subspace associated to the atoms of the molecule.

The transmission function is given by the following version of the
Landauer formula\cite{meir92}
\begin{equation}
\label{e18}
T(E) = \text{Tr}~\{\Gamma_\text{L} G \Gamma_\text{R} G^\dagger\}
\end{equation}
with $\Gamma_{\text{L,R}}$  defined by
\begin{equation}
\Gamma_{\text{L,R}} = \mathfrak{i}(\Sigma_{\text{L,R}} - \Sigma_{\text{L,R}}^\dagger).
\label{EGamma}
\end{equation}

\section{LDOS projected on molecular orbitals}
\label{AppC}

In order to illustrate more clearly  the mixing of molecular states,
we here introduce a projection of the LDOS 
on specific ``coupled molecular orbitals'', $|\mu\rangle$. 
The formal definition of such  orbitals is given by the following
construction. Again, we construct the KS-Hamiltonian $H_\text{KS}$
from the KS-orbitals and energies of the DFT-calculation that has been
done  for the extended molecule (C$_{60}$ plus parts of the leads). 
In the orthogonalized atomic basis set one can partition $H_\text{KS}$
in the following way, 
\begin{equation}
H_{\text{KS}} = \begin{pmatrix}    	H_\text{L}                & V_{\text{LC}}           & V_{\text{LR}}  \\
					V_{\text{LC}}^{*}         & H_\text{C}              & V_{\text{RC}}  \\
					V_{\text{LR}}^{*}         & V_{\text{RC}}^{*}       & H_\text{R}     \\\end{pmatrix}.
\label{EH}
\end{equation}
where as usual $L,R$ refer to the Hilbert spaces of the left and right
electrode and $C$ comprises the remaining part of the full Hilbert
space that belongs to the molecule. 
The states $\mu\rangle$ are the eigenvectors of the  central block
$H_\text{C}$.  They are   
related to the molecular states of C$_{60}$ in vacuum, but some 
effects of the electrode coupling are taken into account. Since the
states ${\mid}\mu \rangle$ form a complete basis of the molecular 
Hilbert-subspace corresponding to  $H_\text{C}$, 
the LDOS on the molecule can then be decomposed into the contributions of each molecular orbital
\begin{equation}
\text{LDOS}(E) = -\frac{1}{\pi} \sum_{\mu}{\langle}\mu {\mid}\Im~G(E){\mid}\mu\rangle.
\end{equation}
The LDOS projected on a molecular orbital ${\mid} \mu \rangle$ can then be identified as 
$-\frac{1}{\pi}{\langle}\mu{\mid}\Im~G(E){\mid}\mu\rangle$.

\begin{figure}[tbp]
\begin{center}
\includegraphics[width=1\columnwidth]{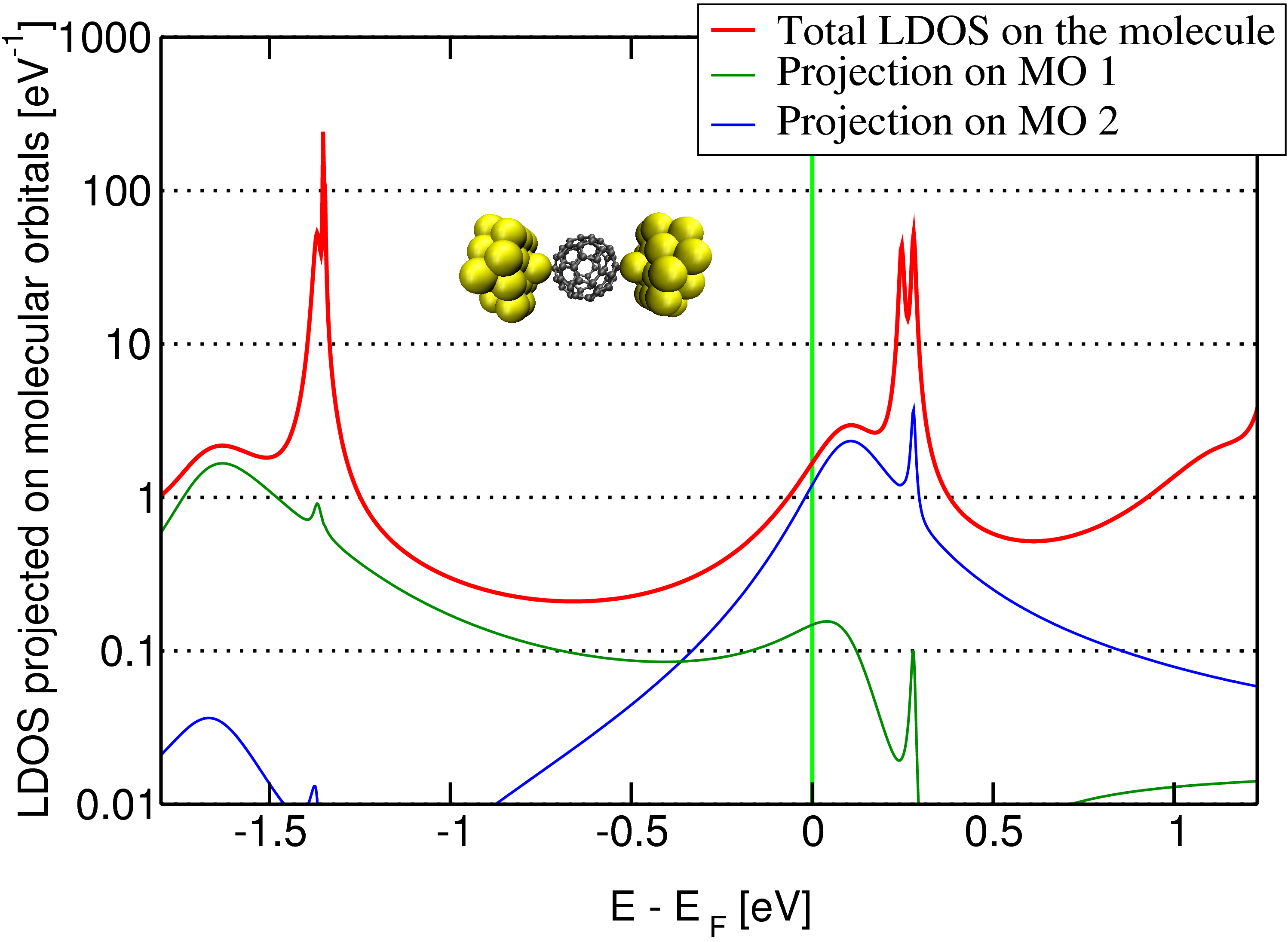}
\end{center}
\caption{(Color) Local density of states projected on the two strongly coupled molecular orbitals (MO 1 and MO 2) for the geometry where 
the adatom sits on-top of a C-atom.}
\label{FTEProjected}
\end{figure}

The local  density of states projected on the two relevant molecular orbitals is
depicted in Fig. \ref{FTEProjected}. It shows that in the valley region of the transmission
between HOMO- and LUMO-resonances two associated orbitals contribute 
similarly to the LDOS. This suggests that there
is a possibility for these orbitals to give interfering
terms in  the transmission function $T(E)$. 

%%%%%%%%%%%%%%%%%%%%%%%%%%%%%%%%%%%%%%%%%%%%%%%%%%%%%%%%%%%%%%%%%%%%%%%%%%%%%%%%%

%%%%%%%%%%%%%%%%%%%%%%%%%%%%%%%%%%%%%%%%%%%%%%%%%%%%%%%%%%%%%%%%%%%%%%%%%%%%%%%%%

\begin{figure}[tbp]
\begin{center}
\includegraphics[width=1\columnwidth]{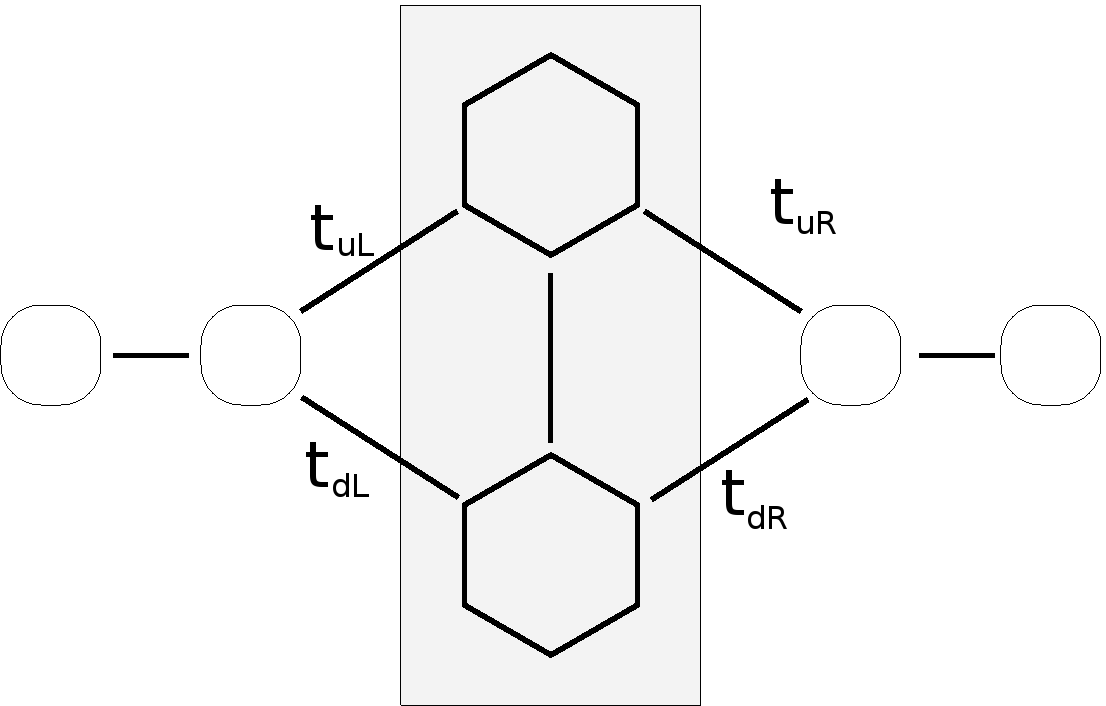}
\end{center}
\caption{Schematic representation of the two-level model with the
inner part representing the Hamiltonian $H$ (grey shaded) and the
couplings to the leads.}
\label{TwoLevelSchematic}
\end{figure}

%%%%%%%%%%%%%%%%%%%%%%%%%%%%%%%%%%%%%%%%%%%%%%%%%%%%%%
\section{Two level (toy) model}
\label{AppD}
%%%%%%%%%%%%%%%%%%%%%%%%%%%%%%%%%%%%%%%%%%%%%%%%%%%%%%

%%%%%%%%%%%%%%%%%%%%%%%%%%%%%%%%%%%%%%%%%%%%%%%%%%%%%%
We recall properties of the two-state (toy) model that accounts for
the transport characteristics of non-interacting quantum dots
with two effective transport levels. As opposed to earlier
work\cite{solomon08a},  we
investigate the model analytically in its full parameter space.

%%%%%%%%%%%%%%%%%%%%%%%%%%%%%%%%%%%%%%%%%%%%%%%%%%%%%%
\subsection{Definition}
Quite generally, the transmission is given by a formula of the Landauer type
(\ref{e18}), 
%\begin{equation}
$
T(E) = \text{Tr}~\{\Gamma_{\text{L}} G \Gamma_{\text{R}} G^\dagger\}.
$
%\end{equation}
For toy model $\Gamma_\text{L,R}$ and $G$ are $2\times
 2$-matrices. We have for the inverse Green's function
\be
G^{-1}(E) = E - H -\Sigma(E)
\ee
where $H$ is a non-interacting two-level Hamiltonian and
$\Sigma(E)$ denotes the self-energy that describes the coupling to
a left and right single chanel wire:
$\Sigma=\Sigma_\text{R}+\Sigma_\text{L}$.
This self energy has the general structure
\bea
\Sigma_\alpha(E) &=& g_\alpha(E)
\left(
\begin{array}{cc}
  |t_{u\alpha}|^2 & t^*_{u\alpha}t_{d\alpha} \\
  t^*_{d\alpha}t_{u\alpha} & |t_{d\alpha}|^2
  \end{array}
\right)\\
&=&
\left(
\begin{array}{c}
  t^*_{u\alpha}\\
  t^*_{d\alpha}
  \end{array}
\right)\ 
g_\alpha(E)\ 
(t_{u\alpha}, t_{d\alpha}), \quad \alpha{=}\text{L,R} 
\label{eD3}
\eea
where $t_{u\alpha},t_{d\alpha}$ denote the hybridization matrix
elements that connect the two level, up and down, with the
wire reservoirs; $g_\alpha(E)$ resembles a scalar, complex
valued function, the "surface Greensfunction" of each wire 
at the point contacting to the two-level system. We have
\be
\Gamma_\alpha = -\ci \left(\Sigma_\alpha(E) - \Sigma_\alpha^\dagger(E)\right);
\ee
Interference effects can occur, if $\Gamma_\text{L,R}$ and $G$
do not commute, so that they cannot be diagonalized simultaneously. 
In order to highlight them, we begin
substituting (\ref{eD3}) into the trace formula; we obtain
\be
T(E) =
(2\pi)^2 \varrho_\text{L} \varrho_\text{R}
\left| ( t_{u\text{L}}t_{d\text{L}}) G
\left(
\begin{array}{c}
  t^*_{u\text{R}}\\
  t^*_{d\text{R}}
  \end{array}
\right)
\right|^2. 
\ee
with the contact density of states
$\varrho(E) = \frac{-1}{\pi}\Im g_\alpha(E)$. 
Next, we rotate into the basis of eigenfunctions
of $G(E)$ 
\begin{equation}
G(E) =  U
\begin{pmatrix}
  \frac{1}{E - z_0}	&	0  \\
0 & \frac{1}{E - z_1}		\\
\end{pmatrix} U^{-1}
\end{equation}
In general, the pole positions $z_{0,1}$ and the eigenvectors
comprising the columns of the rotation matrix
$U$ inherit a dependency on energy, $E$, through $\Sigma(E)$. 
%The situation simplifies in the
%broad-band limit where $\rho_{\alpha}$ and as a
%consequence also $\Gamma,U,z_{0,1}$ become energy independent. 
In order to simplify notation
we introduce effective hybridization matrix elements,
$v$
\bea
\label{e9}
(v_{0\text{L}},v_{1\text{L}}) &=&
\sqrt{2\pi \varrho_\alpha}
(t_{u\text{L}}t_{d\text{L}})U \\
\left(
\begin{array}{c}
v^*_{0\text{R}} \\
v^*_{1\text{R}}
\end{array}
\right)
&=&
\label{e10}
\sqrt{2\pi \varrho_\alpha}
U^{-1}
\left(
\begin{array}{c}
  t^*_{u\text{R}}\\
  t^*_{d\text{R}}
  \end{array}
\right)
\eea
and transmission coefficients
\bea
\tau_{\alpha\beta}(E) &=&
2\pi \sqrt{\varrho_\alpha\varrho_\beta}
( t_{u\alpha}t_{d\alpha}) G
\left(
\begin{array}{c}
  t^*_{u\beta}\\
  t^*_{d\beta}
  \end{array}
\right) \nonumber\\
&=& \frac{v_{0\alpha}v^*_{0\beta}}{E-z_0}
+ \frac{v_{1\alpha}v^*_{1\beta}}{E-z_1}
\eea
that allow us to write
\be
T(E) = |\tau_{\text{LR}}(E)|^2
\ee
To explicitly single out the interference term,
we employ a decomposition
\begin{equation}
T(E) = T_0(E) + T_1(E) + T_{01}(E).
\label{ETEExplicit}
\end{equation}
The two first terms constitute the non-mixing contributions from each energy level
%\begin{eqnarray}
  \be
T_0(E) = \frac{|v_{0\text{L}}|^2 |v_{0\text{R}}|^2 }{|E-z_0|^2},
\quad 
T_1(E) = \frac{|v_{1\text{L}}|^2 |v_{1\text{R}}|^2 }{|E-z_1|^2}
\label{e14}
\ee
%\end{eqnarray}
Each term refers to a single pole only and thus is reproduced
by the model of isolated resonances, Eq. (\ref{e2}). 
Interference enters via the mixed term
\bea
\label{eD13}
T_{01}(E) &=& 2 \sqrt{T_0 T_1} \cos{\left(\Psi_c - \Theta \right)} \\
\Psi_c
&=&\arg(v_{0\text{L}}v^*_{0\text{R}}v^*_{1\text{L}}v_{1\text{R}}) \\
\Theta(E) &=& \arg((E-z_0)(E-z^*_1))
\label{eD15}
\eea
As is seen from equation (\ref{eD13}),
interference effects are controlled by two angles,
$\Psi_c$ and $\Theta$. They exhibit quite different generic
properties. $\Psi_c$ carries an energy dependency only via
$\Sigma(E)$ that reflects a dispersion in the (local) density of
states in the reservoirs. Because the latter often is very smooth
compared to the level splitting, $|\epsilon_0-\epsilon_1|$ with
$\Re z_i = \epsilon_i$, it can typically
be ignored, so that for practical purposes $\Psi_c$ is energy
independent.
By contrast, $\Theta(E)$ can exhibit a very sharp energy
dependency, especially in the limit of low damping. 

%%%%%%%%%%%%%%%%%%%%%%%%%%%%%%%%%%%%%%%%%%%%%%%%%%%%%%%%%%%%
\subsubsection*{Remark: Level broadening}

To reveal further transport properties of the two-level transmission function,
we relate the amplitudes $|v_{0\alpha}|^2$ to the level broadenings
$\gamma_{i}$. This broadening is originates from the anti-hermitian part of the
inverse Green's function
\bea
\label{eD16}
G^{-1} &=& G^{-1}_{0} + \frac{1}{2\ci}\Gamma \\
G^{-1}_{0} &=& E - H - \frac{1}{2}\left( \Sigma + \Sigma^\dagger\right).  
\eea
We have
\be
\text{Tr}G^{-1} = 2E-z_0-z_1 = \text{Tr}\ G^{-1}_{0}
+ \frac{1}{2\ci}\text{Tr}\ \Gamma. 
\ee
Since $G^{-1}_{0},\Gamma$ by construction are hermitian,
each trace is real; hence
\bea
\Re[z_0+z_1] &=& \text{Tr}\ \left[H
+ \frac{1}{2}\left( \Sigma+\Sigma^\dagger\right) \right]\\
\Im[z_0+z_1] &=&  \frac{1}{2}\text{Tr}\ \Gamma. 
\eea
The second line simplifies after recalling (\ref{eD3}): 
\bea
\Im[z_0+z_1] &=&  \frac{1}{2}\left( -2\pi \sum_{\alpha}
\rho_{\alpha}(|t_{u\alpha}|^2 + |t_{d\alpha}|^2) \right) \\
&=& - \frac{1}{2} \sum_{\alpha} |v_{0\alpha}|^2 + |v_{1\alpha}|^2
\eea

%%%%%%%%%%%%%%%%%%%%%%%%%%%%%%%%%%%%%%%%%%%%%%%%%%%%%%%%%%%%
\subsubsection*{Remark: Unitarity theorem}
We split the transmission coeffients into
hermitian and anti-hermitian contributions:
\bea
\tau_{\alpha\beta} &=&
2\pi \sqrt{\varrho_\alpha\varrho_\beta}
( t_{u\alpha}t_{d\alpha}) G
\left(
\begin{array}{c}
  t^*_{u\beta}\\
  t^*_{d\beta}
  \end{array}
\right) \nonumber \\
&=& 2\pi \sqrt{\varrho_\alpha\varrho_\beta}
( t_{u\alpha}t_{d\alpha})
\left[
\frac{1}{2}(G{+}G^\dagger)
{+} \frac{\ci}{2} G\Gamma G^\dagger
\right]
\left(
\begin{array}{c}
  t^*_{u\beta}\\
  t^*_{d\beta}
  \end{array}
\right) \nonumber.
\eea
The first term simplifies to 
$
\frac{1}{2}\left[
\tau_{\alpha\beta} +
\tau^*_{\beta\alpha}
\right]
$
while the second one takes the form 
\bea
&&\frac{(2\pi)^2}{2\ci} \sqrt{\varrho_\alpha\varrho_\beta}
( t_{u\alpha}t_{d\alpha}) \cdot \nonumber\\
&&\hspace{2em} \cdot~ G
\left[
\sum_{\bar \alpha}
\varrho_{\bar\alpha}
\left(
\begin{array}{c}
  t^*_{u\bar\alpha}\\
  t^*_{d\bar\alpha}
  \end{array}
\right) 
( t_{u\bar\alpha}t_{d\bar\alpha})
\right]
G^\dagger
\left(
\begin{array}{c}
  t^*_{u\beta}\\
  t^*_{d\beta}
  \end{array}
\right) \nonumber\\
& &
= \frac{1}{2\ci}
\sum_{\bar\alpha}\tau_{\alpha\bar\alpha}\tau^*_{\beta\bar\alpha} 
%[\tau\tau^\dagger]_{\alpha\beta}
\eea
Summarizing, we have in a matrix notation the general statement 
\be
\tau-\tau^\dagger = -\ci \tau \tau^{\dagger}
\label{eD28}
\ee
that satisfies a unitarity theorem
$\tau^{-1}-[\tau^\dagger]^{-1}=\ci$.

%%%%%%%%%%%%%%%%%%%%%%%%%%%%%%%%%%%%%%%%%%%%%%%%%%%%%%%%%%%
\subsection{Time reversal symmetry}
%%%%%%%%%%%%%%%%%%%%%%%%%%%%%%%%%%%%%%%%%%%%%%%%%%%%%%%%%%%

In the presence of time reversal symmetry the matrices
$G^{-1}_0$ and $\Gamma$ are both real symmetric, so
$G^{-1}$ is (complex) symmetric and
$U$ is (complex) orthogonal. Then
simplifications arise for $T_{01}$.

%%%%%%%%%%%%%%%%%%%%%%%%%%%%%%%%%%%%%%%%%%%%%%%%%%%%%%
We focus the discussion on the weak coupling limit, where
the resonance positions 
are split by an amount that considerably exceeds their
broadening, $|\epsilon_{1}-\epsilon_{0}|\gg \gamma_{0,1}$.
In this case we can consider the
anti-hermitian piece of the self energy $\frac{-\ci}{2}\Gamma$
as a perturbation and the eigenvectors ${\bf u}_{i}$
are real to first order in $\gamma_{i}$;
at least to this order $U$ is real orthogonal.
Hence, also the effective
hybridization matrix elements $v_{i\alpha}$ are real and therefore
$\Psi_c$ takes values zero or $\pi$; we have
\be
\cos(\Psi_c-\Theta(E)) =
\text{sign}(v_{0\text{L}}v_{0\text{R}}v_{1\text{L}}v_{1\text{R}})
     \ \cos\Theta(E)
     \label{e21}
\ee
To the leading order in $\gamma_{i}$ the energy dependency is
described by
\be
\cos\Theta(E) = \text{sign}\left( [E{-}\epsilon_0][E{-}\epsilon_1]\right).
     \label{eD21}
\ee

Equations (\ref{e21},\ref{e21}) are
important because they highlight two
characteristic features of the interference term $T_{01}$.
First, whether constructive ($\cos(\Psi_c-\Theta)> 0$)
or destructive ($\cos(\Psi_c-\Theta) <0$) interference prevails
in the intermediate range $\epsilon_0<E<\epsilon_1$
is controlled by the model dependent first factor
in (\ref{e21}). If it is positive, $\Psi_c=0$,
interference is destructive; it is constructive
in the alternative case $\Psi_c=\pi$. 

Second, $T_{01}$
has a very sharp dependency on energy when
$E$ sweeps by $\epsilon_0$ or $\epsilon_{1}$.
In fact, the corresponding derivative diverges in the limit
$\gamma_{i}\to 0$; specifically, with Eq. (\ref{eD15})
we have 
\bea
\label{eD22}
\cos\Theta(E)
&=& \frac{(E-\epsilon_0)(E-\epsilon_1)+\gamma_0\gamma_1}{|E-z_0| |E-z_1|}\\
\sin\Theta(E) &=&
\frac{\gamma_1(E-\epsilon_0) +\gamma_0(E-\epsilon_1)}{|E-z_0||E-z_1|}
\eea

%%%%%%%%%%%%%%%%%%%%%%%%%%%%%%%%%%%%%%%%%%%%%%%%%%%%%%
\subsection{Limiting cases: symmetric vs asymmetric lead couplings}
%%%%%%%%%%%%%%%%%%%%%%%%%%%%%%%%%%%%%%%%%%%%%%%%%%%%%%%%%%%

We now analyze two limiting cases in which the angle 
$\Psi_c$ is either close to $\pi$ (constructive interference) or
close to zero (destructive interference).

%%%%%%%%%%%%%%%%%%%%%%%%%%%%%%%%%%%%%%%%%%%%%%%%%%%%%%%%%%%
\subsubsection{Fully symmetric coupling: Fano anti-resonance}

We begin with the second
situation, destructive interference,
which is the easier one to investigate
and exploit Eq. (\ref{eD28}) now for
a system with completely symmetric coupling,
$t_{u\text{L}}=t_{u\text{R}}=t_u$, $t_{d\text{L}}=t_{d\text{R}}=t_d$.
This situation is realized, e.g., to a very good approximation 
for cross-conjugated molecular wires with side coupling 
chains as investigated in Ref. [\onlinecite{solomon08b}]. 
A cross-conjugated molecule was studied recently 
experimentally in Ref. [\onlinecite{guedon12}], 
where the Fano-antiresonance has indeed been observed. 

Our interest is in the off-diagonal matrix element, 
\bea
\tau_{\text{L}\text{R}}(E) &=&
\Re[\tau_{\text{LR}}(E)] - \frac{\ci}{2}
\frac{(\rho_\text{L}{+}\rho_\text{R})}
{\sqrt{\varrho_\text{L}\varrho_\text{R}}}
|\tau_{\text{LR}}(E)|^2.
\label{eD27}
\eea
This equation implies the following fact:
Let $E^*$ be a root of
the real part of the transmission coefficients,
$\Re[\tau_{\text{LR}}(E^*)]=0$. Then, at this energy also the imaginary
piece has a vanishing
physical solution: $\Im[\tau_{\text{LR}}(E^*)]=0$.

For generic situations $E^*$ can be
shown to be real, so that there is
no transmission at this
energy, $T(E^*)=0$. Namely, let $\tilde U(E)$ be the unitary rotation
diagonalizing the hermitian matrix $G+G^\dagger$; then 
\be
\Re[\tau_{\text{LR}}(E)] = 2\pi\sqrt{\rho_\text{L}\rho_\text{R}}\left[ \frac{|\tilde v_0|^2}{E-\tilde \epsilon_0}
+ \frac{|\tilde v_1|^2}{E-\tilde \epsilon_1}\right]
\label{eD30}
\ee
where $\tilde\epsilon_{0,1}$ denote the real eigenvalues of
$(G+G^\dagger)/2$ and
$
(\tilde v_0,\tilde v_1) = (t_u,t_d)\tilde U(E).
$
Under the assumption, that the energy dependency of
the eigenvalues is weak (i.e. in the wide band limit, where
$\varrho_{\alpha}(E)$ is nearly constant),
Eq. (\ref{eD30}) has the real numbered root
\be
E^* = \frac{\tilde\epsilon_0|\tilde v_1|^2
+ \tilde\epsilon_1 |\tilde v_0|^2}{|\tilde v_0|^2 + |\tilde v_1|^2}.
\ee It constitutes a weighed average that is situated between the two
pole positions. 

Since at $E^*$ we have complete destructive interference, it is
clear that at this energy $\Psi_c-\Theta(E^*)=\pi$. Since in the
valley region $\Theta$ is close to $\pi$ we have
$\Phi_c\approx 0$ with corrections in $\gamma_{0,1}$.
Therefore, in the valley region away from the anti-resonance $E^*$
we expect 
\be
T_{01}=2\sqrt{T_0T_1}\cos\Theta(E)
\ee
with $\cos\Theta(E)$ as given in Eq. (\ref{eD22}). 

%%%%%%%%%%%%%%%%%%%%%%%%%%%%%%%%%%%%%%%%%%%%%%%%%%%%%%%%%%%
\subsubsection{Fully asymetric coupling}
 
The previous example suggests, that constructive interference could be
expected for an asymetric limit, $t_{u\text{R}},t_{d\text{L}}=0$.
In this case,
because of equations (\ref{e9},\ref{e10}) the effective
hybridization matrix elements reproduce the
entries of $U$ and $U^{-1}$:
\bea
v_{i\text{L}}&=&\sqrt{2\pi\varrho_{\text{L}}}t_{u\text{L}}U_{0i} \nonumber\\
v^*_{i\text{R}}&=&\sqrt{2\pi\varrho_{\text{R}}}t^*_{d\text{R}}[U^{-1}]_{i1} \nonumber
\eea
so that
\be
%%                           v0L      v0R*             v1R    v1L*
\Psi_c = \arg\left( U_{00} [U^{-1}]_{01} [U^{-1}]^*_{11}  U^*_{01} \right) 
\label{eD26}
\ee
The product in brackets is easy to evaluate in the presence
of time reversal symmetry 
recalling that $U$ is orthogonal in this case.
Like any orthogonal $2\times 2$-rotation
matrix it has a representation
\be
  U =
  \left(
  \begin{array}{cc}
    \cos w  & \sin w \\
    -\sin w  & \cos w
    \end{array}
  \right),\qquad w\in \mathbb{C}
  \nonumber
%  }
\ee
We obtain $\Psi_c=\pi$. Hence, we now expect a 
constructive interference contribution
\be
T_{01}= - 2\sqrt{T_0T_1}\cos\Theta(E). 
\ee
where again $\cos\Theta(E)$ is given in Eq. (\ref{eD22}).

%%%%%%%%%%%%%%%%%%%%%%%%%%%%%%%%%%%%%%%%%%%%%%%%%%%%%%
%%
%%%%%%%%%%%%%%%%%%%%%%%%%%%%%%%%%%%%%%%%%%%%%%%%%%%%%%

%\bibliographystyle{unsrt}
\bibliographystyle{apsrev}
%\bibliography{./BibME/bibDft,../BibME/bibExperiments,./BibME/bibDftApplTransport,./BibME/bibInelastic,./BibME/bibOwnMolEl,./BibME/bibNegf,./BibME/bibTDFT,./BibME/books,./BibME/bibGeneral,./BibME/bibMethods}
\bibliography{./Biblio,../../BibME/bibExperiments,../../BibME/bibInelastic,../../BibME/bibDft,../../BibME/bibNegf}

\begin{thebibliography}{55}
\expandafter\ifx\csname natexlab\endcsname\relax\def\natexlab#1{#1}\fi
\expandafter\ifx\csname bibnamefont\endcsname\relax
  \def\bibnamefont#1{#1}\fi
\expandafter\ifx\csname bibfnamefont\endcsname\relax
  \def\bibfnamefont#1{#1}\fi
\expandafter\ifx\csname citenamefont\endcsname\relax
  \def\citenamefont#1{#1}\fi
\expandafter\ifx\csname url\endcsname\relax
  \def\url#1{\texttt{#1}}\fi
\expandafter\ifx\csname urlprefix\endcsname\relax\def\urlprefix{URL }\fi
\providecommand{\bibinfo}[2]{#2}
\providecommand{\eprint}[2][]{\url{#2}}

\bibitem[{\citenamefont{Lu et~al.}(2003)\citenamefont{Lu, Grobis, Khoo, Louie,
  and Crommie}}]{lu03}
\bibinfo{author}{\bibfnamefont{X.}~\bibnamefont{Lu}},
  \bibinfo{author}{\bibfnamefont{M.}~\bibnamefont{Grobis}},
  \bibinfo{author}{\bibfnamefont{K.~H.} \bibnamefont{Khoo}},
  \bibinfo{author}{\bibfnamefont{S.~G.} \bibnamefont{Louie}}, \bibnamefont{and}
  \bibinfo{author}{\bibfnamefont{M.~F.} \bibnamefont{Crommie}},
  \bibinfo{journal}{Phys. Rev. Lett.} \textbf{\bibinfo{volume}{90}},
  \bibinfo{pages}{096802} (\bibinfo{year}{2003}),
  \urlprefix\url{http://link.aps.org/doi/10.1103/PhysRevLett.90.096802}.

\bibitem[{\citenamefont{Lu et~al.}(2004)\citenamefont{Lu, Grobis, Khoo, Louie,
  and Crommie}}]{lu04}
\bibinfo{author}{\bibfnamefont{X.}~\bibnamefont{Lu}},
  \bibinfo{author}{\bibfnamefont{M.}~\bibnamefont{Grobis}},
  \bibinfo{author}{\bibfnamefont{K.~H.} \bibnamefont{Khoo}},
  \bibinfo{author}{\bibfnamefont{S.~G.} \bibnamefont{Louie}}, \bibnamefont{and}
  \bibinfo{author}{\bibfnamefont{M.~F.} \bibnamefont{Crommie}},
  \bibinfo{journal}{Phys. Rev. B} \textbf{\bibinfo{volume}{70}},
  \bibinfo{pages}{115418} (\bibinfo{year}{2004}).

\bibitem[{\citenamefont{Rogero et~al.}(2002)\citenamefont{Rogero, Pascual,
  G\'omez-Herrero, and Bar\'o}}]{rogero02}
\bibinfo{author}{\bibfnamefont{C.}~\bibnamefont{Rogero}},
  \bibinfo{author}{\bibfnamefont{J.~I.} \bibnamefont{Pascual}},
  \bibinfo{author}{\bibfnamefont{J.}~\bibnamefont{G\'omez-Herrero}},
  \bibnamefont{and} \bibinfo{author}{\bibfnamefont{A.~M.}
  \bibnamefont{Bar\'o}}, \bibinfo{journal}{J. Chem. Phys.}
  \textbf{\bibinfo{volume}{116}}, \bibinfo{pages}{832} (\bibinfo{year}{2002}).

\bibitem[{\citenamefont{Sau et~al.}(2008)\citenamefont{Sau, Neaton, Choi,
  Louie, and Cohen}}]{sau08}
\bibinfo{author}{\bibfnamefont{J.~D.} \bibnamefont{Sau}},
  \bibinfo{author}{\bibfnamefont{J.~B.} \bibnamefont{Neaton}},
  \bibinfo{author}{\bibfnamefont{H.~J.} \bibnamefont{Choi}},
  \bibinfo{author}{\bibfnamefont{S.~G.} \bibnamefont{Louie}}, \bibnamefont{and}
  \bibinfo{author}{\bibfnamefont{M.~L.} \bibnamefont{Cohen}},
  \bibinfo{journal}{Phys. Rev. Lett.} \textbf{\bibinfo{volume}{101}},
  \bibinfo{pages}{026804} (\bibinfo{year}{2008}),
  \urlprefix\url{http://link.aps.org/doi/10.1103/PhysRevLett.101.026804}.

\bibitem[{\citenamefont{Torrente et~al.}(2008)\citenamefont{Torrente, Franke,
  and Pascual}}]{torrente08}
\bibinfo{author}{\bibfnamefont{I.~F.} \bibnamefont{Torrente}},
  \bibinfo{author}{\bibfnamefont{K.~J.} \bibnamefont{Franke}},
  \bibnamefont{and} \bibinfo{author}{\bibfnamefont{J.~I.}
  \bibnamefont{Pascual}}, \bibinfo{journal}{J. Phys.: Condens. Matter}
  \textbf{\bibinfo{volume}{20}}, \bibinfo{pages}{184001}
  (\bibinfo{year}{2008}).

\bibitem[{\citenamefont{Tamai et~al.}(2008)\citenamefont{Tamai, Seitsonen,
  Baumberger, Hengsberger, Shen, Greber, and Osterwalder}}]{tamai08}
\bibinfo{author}{\bibfnamefont{A.}~\bibnamefont{Tamai}},
  \bibinfo{author}{\bibfnamefont{A.~P.} \bibnamefont{Seitsonen}},
  \bibinfo{author}{\bibfnamefont{F.}~\bibnamefont{Baumberger}},
  \bibinfo{author}{\bibfnamefont{M.}~\bibnamefont{Hengsberger}},
  \bibinfo{author}{\bibfnamefont{Z.-X.} \bibnamefont{Shen}},
  \bibinfo{author}{\bibfnamefont{T.}~\bibnamefont{Greber}}, \bibnamefont{and}
  \bibinfo{author}{\bibfnamefont{J.}~\bibnamefont{Osterwalder}},
  \bibinfo{journal}{Phys. Rev. B} \textbf{\bibinfo{volume}{77}},
  \bibinfo{pages}{075134} (\bibinfo{year}{2008}).

\bibitem[{\citenamefont{Abad et~al.}(2010)\citenamefont{Abad, Ganzalez, Ortega,
  and Flores}}]{abad10}
\bibinfo{author}{\bibfnamefont{E.}~\bibnamefont{Abad}},
  \bibinfo{author}{\bibfnamefont{C.}~\bibnamefont{Ganzalez}},
  \bibinfo{author}{\bibfnamefont{J.}~\bibnamefont{Ortega}}, \bibnamefont{and}
  \bibinfo{author}{\bibfnamefont{F.}~\bibnamefont{Flores}},
  \bibinfo{journal}{Organic Electronics} \textbf{\bibinfo{volume}{11}},
  \bibinfo{pages}{332} (\bibinfo{year}{2010}).

\bibitem[{\citenamefont{Hamada and Tsukada}(2011)}]{hamada11}
\bibinfo{author}{\bibfnamefont{I.}~\bibnamefont{Hamada}} \bibnamefont{and}
  \bibinfo{author}{\bibfnamefont{M.}~\bibnamefont{Tsukada}},
  \bibinfo{journal}{Phys. Rev. B} \textbf{\bibinfo{volume}{83}},
  \bibinfo{pages}{245437} (\bibinfo{year}{2011}),
  \urlprefix\url{http://link.aps.org/doi/10.1103/PhysRevB.83.245437}.

\bibitem[{\citenamefont{Park et~al.}(2000)\citenamefont{Park, Park, Lim,
  Anderson, Alivisatos, and McEuen}}]{park00}
\bibinfo{author}{\bibfnamefont{H.}~\bibnamefont{Park}},
  \bibinfo{author}{\bibfnamefont{J.}~\bibnamefont{Park}},
  \bibinfo{author}{\bibfnamefont{A.~K.~L.} \bibnamefont{Lim}},
  \bibinfo{author}{\bibfnamefont{E.~H.} \bibnamefont{Anderson}},
  \bibinfo{author}{\bibfnamefont{A.~P.} \bibnamefont{Alivisatos}},
  \bibnamefont{and} \bibinfo{author}{\bibfnamefont{P.~L.}
  \bibnamefont{McEuen}}, \bibinfo{journal}{Nature}
  \textbf{\bibinfo{volume}{407}}, \bibinfo{pages}{57} (\bibinfo{year}{2000}).

\bibitem[{\citenamefont{Yu and Natelson}(2004)}]{yu04}
\bibinfo{author}{\bibfnamefont{L.~H.} \bibnamefont{Yu}} \bibnamefont{and}
  \bibinfo{author}{\bibfnamefont{D.}~\bibnamefont{Natelson}},
  \bibinfo{journal}{Nano Letters} \textbf{\bibinfo{volume}{4}},
  \bibinfo{pages}{79} (\bibinfo{year}{2004}).

\bibitem[{\citenamefont{Pasupathy et~al.}(2004)\citenamefont{Pasupathy,
  Bialczak, Martinek, Grose, Donev, McEuen, and Ralph}}]{pasupathy04}
\bibinfo{author}{\bibfnamefont{A.~N.} \bibnamefont{Pasupathy}},
  \bibinfo{author}{\bibfnamefont{R.~C.} \bibnamefont{Bialczak}},
  \bibinfo{author}{\bibfnamefont{J.}~\bibnamefont{Martinek}},
  \bibinfo{author}{\bibfnamefont{J.~E.} \bibnamefont{Grose}},
  \bibinfo{author}{\bibfnamefont{L.~A.~K.} \bibnamefont{Donev}},
  \bibinfo{author}{\bibfnamefont{P.~L.} \bibnamefont{McEuen}},
  \bibnamefont{and} \bibinfo{author}{\bibfnamefont{D.~C.} \bibnamefont{Ralph}},
  \bibinfo{journal}{Science} \textbf{\bibinfo{volume}{306}},
  \bibinfo{pages}{86} (\bibinfo{year}{2004}).

\bibitem[{\citenamefont{Parks et~al.}(2007)\citenamefont{Parks, Champagne,
  Hutchison, Flores-Torres, Abruna, and Ralph}}]{parks07}
\bibinfo{author}{\bibfnamefont{J.~J.} \bibnamefont{Parks}},
  \bibinfo{author}{\bibfnamefont{A.~R.} \bibnamefont{Champagne}},
  \bibinfo{author}{\bibfnamefont{G.~R.} \bibnamefont{Hutchison}},
  \bibinfo{author}{\bibfnamefont{S.}~\bibnamefont{Flores-Torres}},
  \bibinfo{author}{\bibfnamefont{H.~D.} \bibnamefont{Abruna}},
  \bibnamefont{and} \bibinfo{author}{\bibfnamefont{D.~C.} \bibnamefont{Ralph}},
  \bibinfo{journal}{Phys. Rev. Lett.} \textbf{\bibinfo{volume}{99}},
  \bibinfo{pages}{026601} (\bibinfo{year}{2007}).

\bibitem[{\citenamefont{Roch et~al.}(2008)\citenamefont{Roch, Florens,
  Bouchiat, Wernsdorfer, and Balestro}}]{roch08}
\bibinfo{author}{\bibfnamefont{N.}~\bibnamefont{Roch}},
  \bibinfo{author}{\bibfnamefont{S.}~\bibnamefont{Florens}},
  \bibinfo{author}{\bibfnamefont{V.}~\bibnamefont{Bouchiat}},
  \bibinfo{author}{\bibfnamefont{W.}~\bibnamefont{Wernsdorfer}},
  \bibnamefont{and} \bibinfo{author}{\bibfnamefont{F.}~\bibnamefont{Balestro}},
  \bibinfo{journal}{Nature} \textbf{\bibinfo{volume}{453}},
  \bibinfo{pages}{633} (\bibinfo{year}{2008}).

\bibitem[{\citenamefont{Schulze
  et~al.}(2008{\natexlab{a}})\citenamefont{Schulze, Franke, and
  and}}]{schulze08njp}
\bibinfo{author}{\bibfnamefont{G.}~\bibnamefont{Schulze}},
  \bibinfo{author}{\bibfnamefont{K.~J.} \bibnamefont{Franke}},
  \bibnamefont{and} \bibinfo{author}{\bibfnamefont{J.~I.~P.}
  \bibnamefont{and}}, \bibinfo{journal}{New J. of Phys.}
  \textbf{\bibinfo{volume}{10}}, \bibinfo{pages}{065005}
  (\bibinfo{year}{2008}{\natexlab{a}}).

\bibitem[{\citenamefont{Schulze
  et~al.}(2008{\natexlab{b}})\citenamefont{Schulze, Franke, Gagliardi, Romano,
  Lin, Rosa, Niehaus, and Frauenheim}}]{schulze08prl}
\bibinfo{author}{\bibfnamefont{G.}~\bibnamefont{Schulze}},
  \bibinfo{author}{\bibfnamefont{K.~J.} \bibnamefont{Franke}},
  \bibinfo{author}{\bibfnamefont{A.}~\bibnamefont{Gagliardi}},
  \bibinfo{author}{\bibfnamefont{G.}~\bibnamefont{Romano}},
  \bibinfo{author}{\bibfnamefont{C.~S.} \bibnamefont{Lin}},
  \bibinfo{author}{\bibfnamefont{A.~L.} \bibnamefont{Rosa}},
  \bibinfo{author}{\bibfnamefont{T.~A.} \bibnamefont{Niehaus}},
  \bibnamefont{and}
  \bibinfo{author}{\bibfnamefont{T.}~\bibnamefont{Frauenheim}},
  \bibinfo{journal}{Phys. Rev. Lett.} \textbf{\bibinfo{volume}{100}},
  \bibinfo{pages}{136801} (\bibinfo{year}{2008}{\natexlab{b}}).

\bibitem[{\citenamefont{Frederiksen et~al.}(2008)\citenamefont{Frederiksen,
  Franke, Arnau, Schulze, Pascual, and Lorente}}]{frederiksen08}
\bibinfo{author}{\bibfnamefont{T.}~\bibnamefont{Frederiksen}},
  \bibinfo{author}{\bibfnamefont{K.~J.} \bibnamefont{Franke}},
  \bibinfo{author}{\bibfnamefont{A.}~\bibnamefont{Arnau}},
  \bibinfo{author}{\bibfnamefont{G.}~\bibnamefont{Schulze}},
  \bibinfo{author}{\bibfnamefont{J.~I.} \bibnamefont{Pascual}},
  \bibnamefont{and} \bibinfo{author}{\bibfnamefont{N.}~\bibnamefont{Lorente}},
  \bibinfo{journal}{Phys. Rev. B} \textbf{\bibinfo{volume}{78}},
  \bibinfo{pages}{233401} (\bibinfo{year}{2008}).

\bibitem[{\citenamefont{Gagliardi et~al.}(2008)\citenamefont{Gagliardi, Romano,
  Pecchia, Carlo, Frauenheim, and Nienhaus}}]{gagliardi08}
\bibinfo{author}{\bibfnamefont{A.}~\bibnamefont{Gagliardi}},
  \bibinfo{author}{\bibfnamefont{G.}~\bibnamefont{Romano}},
  \bibinfo{author}{\bibfnamefont{A.}~\bibnamefont{Pecchia}},
  \bibinfo{author}{\bibfnamefont{A.~D.} \bibnamefont{Carlo}},
  \bibinfo{author}{\bibfnamefont{T.}~\bibnamefont{Frauenheim}},
  \bibnamefont{and} \bibinfo{author}{\bibfnamefont{T.~A.}
  \bibnamefont{Nienhaus}}, \bibinfo{journal}{New Journal of Physics}
  \textbf{\bibinfo{volume}{10}}, \bibinfo{pages}{065020}
  (\bibinfo{year}{2008}).

\bibitem[{\citenamefont{Fock et~al.}(2011)\citenamefont{Fock, Sorensen,
  Lortscher, Vosch, Martin, Riel, Kilsa, Bjornholm, and van~der Zant}}]{fock11}
\bibinfo{author}{\bibfnamefont{J.}~\bibnamefont{Fock}},
  \bibinfo{author}{\bibfnamefont{J.~K.} \bibnamefont{Sorensen}},
  \bibinfo{author}{\bibfnamefont{E.}~\bibnamefont{Lortscher}},
  \bibinfo{author}{\bibfnamefont{T.}~\bibnamefont{Vosch}},
  \bibinfo{author}{\bibfnamefont{C.~A.} \bibnamefont{Martin}},
  \bibinfo{author}{\bibfnamefont{H.}~\bibnamefont{Riel}},
  \bibinfo{author}{\bibfnamefont{K.}~\bibnamefont{Kilsa}},
  \bibinfo{author}{\bibfnamefont{T.}~\bibnamefont{Bjornholm}},
  \bibnamefont{and} \bibinfo{author}{\bibfnamefont{H.}~\bibnamefont{van~der
  Zant}}, \bibinfo{journal}{Phys. Chem. Chem. Phys.}
  \textbf{\bibinfo{volume}{13}}, \bibinfo{pages}{14325} (\bibinfo{year}{2011}),
  \urlprefix\url{http://dx.doi.org/10.1039/C1CP20861F}.

\bibitem[{\citenamefont{N\'eel et~al.}(2007)\citenamefont{N\'eel, Kr\"oger,
  Limot, Frederiksen, Brandbyge, and Berndt}}]{neel07}
\bibinfo{author}{\bibfnamefont{N.}~\bibnamefont{N\'eel}},
  \bibinfo{author}{\bibfnamefont{J.}~\bibnamefont{Kr\"oger}},
  \bibinfo{author}{\bibfnamefont{L.}~\bibnamefont{Limot}},
  \bibinfo{author}{\bibfnamefont{T.}~\bibnamefont{Frederiksen}},
  \bibinfo{author}{\bibfnamefont{M.}~\bibnamefont{Brandbyge}},
  \bibnamefont{and} \bibinfo{author}{\bibfnamefont{R.}~\bibnamefont{Berndt}},
  \bibinfo{journal}{Phys. Rev. Lett.} \textbf{\bibinfo{volume}{98}},
  \bibinfo{pages}{065502} (\bibinfo{year}{2007}).

\bibitem[{\citenamefont{Schull et~al.}(2009)\citenamefont{Schull, Frederiksen,
  Brandbyge, and Berndt}}]{schull09}
\bibinfo{author}{\bibfnamefont{G.}~\bibnamefont{Schull}},
  \bibinfo{author}{\bibfnamefont{T.}~\bibnamefont{Frederiksen}},
  \bibinfo{author}{\bibfnamefont{M.}~\bibnamefont{Brandbyge}},
  \bibnamefont{and} \bibinfo{author}{\bibfnamefont{R.}~\bibnamefont{Berndt}},
  \bibinfo{journal}{Phys. Rev. Lett.} \textbf{\bibinfo{volume}{103}},
  \bibinfo{pages}{206803} (\bibinfo{year}{2009}).

\bibitem[{\citenamefont{Schull et~al.}(2010)\citenamefont{Schull, Frederiksen,
  Arnau, Sanchez-Portal, and Berndt}}]{schull10}
\bibinfo{author}{\bibfnamefont{G.}~\bibnamefont{Schull}},
  \bibinfo{author}{\bibfnamefont{T.}~\bibnamefont{Frederiksen}},
  \bibinfo{author}{\bibfnamefont{A.}~\bibnamefont{Arnau}},
  \bibinfo{author}{\bibfnamefont{D.}~\bibnamefont{Sanchez-Portal}},
  \bibnamefont{and} \bibinfo{author}{\bibfnamefont{R.}~\bibnamefont{Berndt}},
  \bibinfo{journal}{Nature Nanotechnology} \textbf{\bibinfo{volume}{6}},
  \bibinfo{pages}{23} (\bibinfo{year}{2010}).

\bibitem[{\citenamefont{Schull et~al.}(2011)\citenamefont{Schull, Dappe,
  Gonzalez, Bulou, and Berndt}}]{schull11}
\bibinfo{author}{\bibfnamefont{G.}~\bibnamefont{Schull}},
  \bibinfo{author}{\bibfnamefont{Y.~J.} \bibnamefont{Dappe}},
  \bibinfo{author}{\bibfnamefont{C.}~\bibnamefont{Gonzalez}},
  \bibinfo{author}{\bibfnamefont{H.}~\bibnamefont{Bulou}}, \bibnamefont{and}
  \bibinfo{author}{\bibfnamefont{R.}~\bibnamefont{Berndt}},
  \bibinfo{journal}{Nano Letters} \textbf{\bibinfo{volume}{11}},
  \bibinfo{pages}{3142} (\bibinfo{year}{2011}).

\bibitem[{\citenamefont{B\"ohler et~al.}(2007)\citenamefont{B\"ohler, Edtbauer,
  and Scheer}}]{boehler07}
\bibinfo{author}{\bibfnamefont{T.}~\bibnamefont{B\"ohler}},
  \bibinfo{author}{\bibfnamefont{A.}~\bibnamefont{Edtbauer}}, \bibnamefont{and}
  \bibinfo{author}{\bibfnamefont{E.}~\bibnamefont{Scheer}},
  \bibinfo{journal}{Phys. Rev. B} \textbf{\bibinfo{volume}{76}},
  \bibinfo{pages}{125432} (\bibinfo{year}{2007}),
  \urlprefix\url{http://link.aps.org/doi/10.1103/PhysRevB.76.125432}.

\bibitem[{\citenamefont{Kiguchi and Murakoshi}(2008)}]{kiguchi08}
\bibinfo{author}{\bibfnamefont{M.}~\bibnamefont{Kiguchi}} \bibnamefont{and}
  \bibinfo{author}{\bibfnamefont{K.}~\bibnamefont{Murakoshi}},
  \bibinfo{journal}{J. Phys. Chem. C} \textbf{\bibinfo{volume}{112}},
  \bibinfo{pages}{8140} (\bibinfo{year}{2008}).

\bibitem[{\citenamefont{Yee et~al.}(2011)\citenamefont{Yee, Malen, Majumdar,
  and Segalman}}]{yee11}
\bibinfo{author}{\bibfnamefont{S.~K.} \bibnamefont{Yee}},
  \bibinfo{author}{\bibfnamefont{J.~A.} \bibnamefont{Malen}},
  \bibinfo{author}{\bibfnamefont{A.}~\bibnamefont{Majumdar}}, \bibnamefont{and}
  \bibinfo{author}{\bibfnamefont{R.~A.} \bibnamefont{Segalman}},
  \bibinfo{journal}{Nano Letters} \textbf{\bibinfo{volume}{11}},
  \bibinfo{pages}{4089} (\bibinfo{year}{2011}).

\bibitem[{\citenamefont{Mayor et~al.}(2003)\citenamefont{Mayor, Weber,
  Reichert, Elbing, von Hanisch, Beckmann, and Fischer}}]{mayor03}
\bibinfo{author}{\bibfnamefont{M.}~\bibnamefont{Mayor}},
  \bibinfo{author}{\bibfnamefont{H.~B.} \bibnamefont{Weber}},
  \bibinfo{author}{\bibfnamefont{J.}~\bibnamefont{Reichert}},
  \bibinfo{author}{\bibfnamefont{M.}~\bibnamefont{Elbing}},
  \bibinfo{author}{\bibfnamefont{C.}~\bibnamefont{von Hanisch}},
  \bibinfo{author}{\bibfnamefont{D.}~\bibnamefont{Beckmann}}, \bibnamefont{and}
  \bibinfo{author}{\bibfnamefont{M.}~\bibnamefont{Fischer}},
  \bibinfo{journal}{Angwandte Chemie-International Edition}
  \textbf{\bibinfo{volume}{2003}}, \bibinfo{pages}{5834}
  (\bibinfo{year}{2003}).

\bibitem[{\citenamefont{Martin et~al.}(2008)\citenamefont{Martin, Ding,
  Sorensen, Bjornholm, van Ruitenbeek, and van~der Zant}}]{martin08}
\bibinfo{author}{\bibfnamefont{C.~A.} \bibnamefont{Martin}},
  \bibinfo{author}{\bibfnamefont{D.}~\bibnamefont{Ding}},
  \bibinfo{author}{\bibfnamefont{J.~K.} \bibnamefont{Sorensen}},
  \bibinfo{author}{\bibfnamefont{T.}~\bibnamefont{Bjornholm}},
  \bibinfo{author}{\bibfnamefont{J.}~\bibnamefont{van Ruitenbeek}},
  \bibnamefont{and} \bibinfo{author}{\bibfnamefont{H.~S.~J.}
  \bibnamefont{van~der Zant}}, \bibinfo{journal}{J. Am. Chem. Soc.}
  \textbf{\bibinfo{volume}{130}}, \bibinfo{pages}{13198}
  (\bibinfo{year}{2008}).

\bibitem[{\citenamefont{Leary et~al.}(2011)\citenamefont{Leary, Gonzalez,
  van~der Pol, Bryce, Filippone, Martin, Rubio-Bollinger, and
  Agrait}}]{leary11}
\bibinfo{author}{\bibfnamefont{E.}~\bibnamefont{Leary}},
  \bibinfo{author}{\bibfnamefont{M.~T.} \bibnamefont{Gonzalez}},
  \bibinfo{author}{\bibfnamefont{C.}~\bibnamefont{van~der Pol}},
  \bibinfo{author}{\bibfnamefont{M.~R.} \bibnamefont{Bryce}},
  \bibinfo{author}{\bibfnamefont{S.}~\bibnamefont{Filippone}},
  \bibinfo{author}{\bibfnamefont{N.}~\bibnamefont{Martin}},
  \bibinfo{author}{\bibfnamefont{G.}~\bibnamefont{Rubio-Bollinger}},
  \bibnamefont{and} \bibinfo{author}{\bibfnamefont{N.}~\bibnamefont{Agrait}},
  \bibinfo{journal}{Nano Letters} \textbf{\bibinfo{volume}{11}},
  \bibinfo{pages}{2236} (\bibinfo{year}{2011}).

\bibitem[{\citenamefont{Arnold et~al.}(2007)\citenamefont{Arnold, Weigend, and
  Evers}}]{TranspForm}
\bibinfo{author}{\bibfnamefont{A.}~\bibnamefont{Arnold}},
  \bibinfo{author}{\bibfnamefont{F.}~\bibnamefont{Weigend}}, \bibnamefont{and}
  \bibinfo{author}{\bibfnamefont{F.}~\bibnamefont{Evers}},
  \bibinfo{journal}{Journal of Chemical Physics}
  \textbf{\bibinfo{volume}{126}}, \bibinfo{pages}{174101}
  (\bibinfo{year}{2007}).

\bibitem[{\citenamefont{Shukla et~al.}(2008)\citenamefont{Shukla, Dubey, and
  Leszczynski}}]{shukla08}
\bibinfo{author}{\bibfnamefont{M.~K.} \bibnamefont{Shukla}},
  \bibinfo{author}{\bibfnamefont{M.}~\bibnamefont{Dubey}}, \bibnamefont{and}
  \bibinfo{author}{\bibnamefont{Leszczynski}}, \bibinfo{journal}{ACS Nano}
  \textbf{\bibinfo{volume}{2}}, \bibinfo{pages}{227} (\bibinfo{year}{2008}).

\bibitem[{\citenamefont{Shukla et~al.}(2009)\citenamefont{Shukla, Dubey, Zakar,
  and Leszczynksi}}]{shukla09}
\bibinfo{author}{\bibfnamefont{M.~K.} \bibnamefont{Shukla}},
  \bibinfo{author}{\bibfnamefont{M.}~\bibnamefont{Dubey}},
  \bibinfo{author}{\bibfnamefont{E.}~\bibnamefont{Zakar}}, \bibnamefont{and}
  \bibinfo{author}{\bibfnamefont{J.}~\bibnamefont{Leszczynksi}},
  \bibinfo{journal}{J. Phys. Chem.} \textbf{\bibinfo{volume}{113}},
  \bibinfo{pages}{11351} (\bibinfo{year}{2009}).

\bibitem[{\citenamefont{Bilan et~al.}(2012)\citenamefont{Bilan, Zotti, Pauly,
  and Cuevas}}]{bilan12}
\bibinfo{author}{\bibfnamefont{S.}~\bibnamefont{Bilan}},
  \bibinfo{author}{\bibfnamefont{L.~A.} \bibnamefont{Zotti}},
  \bibinfo{author}{\bibfnamefont{F.}~\bibnamefont{Pauly}}, \bibnamefont{and}
  \bibinfo{author}{\bibfnamefont{J.~C.} \bibnamefont{Cuevas}},
  \bibinfo{journal}{arXiv:1203.3101v1}  (\bibinfo{year}{2012}).

\bibitem[{\citenamefont{Ono and Hirose}(2007)}]{ono07}
\bibinfo{author}{\bibfnamefont{T.}~\bibnamefont{Ono}} \bibnamefont{and}
  \bibinfo{author}{\bibfnamefont{K.}~\bibnamefont{Hirose}},
  \bibinfo{journal}{Phys. Rev. Lett.} \textbf{\bibinfo{volume}{98}},
  \bibinfo{pages}{026804} (\bibinfo{year}{2007}).

\bibitem[{\citenamefont{Ahlrichs et~al.}(1989)\citenamefont{Ahlrichs, B{\"a}r,
  H{\"a}ser, Horn, and K{\"o}lmel}}]{turbomole89}
\bibinfo{author}{\bibfnamefont{R.}~\bibnamefont{Ahlrichs}},
  \bibinfo{author}{\bibfnamefont{M.}~\bibnamefont{B{\"a}r}},
  \bibinfo{author}{\bibfnamefont{M.}~\bibnamefont{H{\"a}ser}},
  \bibinfo{author}{\bibfnamefont{H.}~\bibnamefont{Horn}}, \bibnamefont{and}
  \bibinfo{author}{\bibfnamefont{C.}~\bibnamefont{K{\"o}lmel}},
  \bibinfo{journal}{Chem. Phys. Lett.} \textbf{\bibinfo{volume}{162}},
  \bibinfo{pages}{165} (\bibinfo{year}{1989}).

\bibitem[{\citenamefont{Becke}(1988)}]{becke88}
\bibinfo{author}{\bibfnamefont{A.}~\bibnamefont{Becke}},
  \bibinfo{journal}{Phys. Rev. A} \textbf{\bibinfo{volume}{38}},
  \bibinfo{pages}{3098} (\bibinfo{year}{1988}).

\bibitem[{\citenamefont{Grimme}(2004)}]{Grimme2004}
\bibinfo{author}{\bibfnamefont{S.}~\bibnamefont{Grimme}},
  \bibinfo{journal}{Journal of Computational Chemistry}
  \textbf{\bibinfo{volume}{25}}, \bibinfo{pages}{1463} (\bibinfo{year}{2004}).

\bibitem[{\citenamefont{Schneebeli et~al.}(2011)\citenamefont{Schneebeli,
  Kamenetska, Cheng, Skouta, Friesner, Venkataraman, and
  Breslow}}]{schneebeli11}
\bibinfo{author}{\bibfnamefont{S.~T.} \bibnamefont{Schneebeli}},
  \bibinfo{author}{\bibfnamefont{M.}~\bibnamefont{Kamenetska}},
  \bibinfo{author}{\bibfnamefont{Z.}~\bibnamefont{Cheng}},
  \bibinfo{author}{\bibfnamefont{R.}~\bibnamefont{Skouta}},
  \bibinfo{author}{\bibfnamefont{R.~A.} \bibnamefont{Friesner}},
  \bibinfo{author}{\bibfnamefont{L.}~\bibnamefont{Venkataraman}},
  \bibnamefont{and} \bibinfo{author}{\bibfnamefont{R.}~\bibnamefont{Breslow}},
  \bibinfo{journal}{J. Am. Soc. Chem.} \textbf{\bibinfo{volume}{133}},
  \bibinfo{pages}{2136} (\bibinfo{year}{2011}).

\bibitem[{\citenamefont{Lof et~al.}(1992)\citenamefont{Lof, van Veenendaal,
  Koopmans, Jonkman, and Sawatzky}}]{GapC60}
\bibinfo{author}{\bibfnamefont{R.~W.} \bibnamefont{Lof}},
  \bibinfo{author}{\bibfnamefont{M.~A.} \bibnamefont{van Veenendaal}},
  \bibinfo{author}{\bibfnamefont{B.}~\bibnamefont{Koopmans}},
  \bibinfo{author}{\bibfnamefont{H.~T.} \bibnamefont{Jonkman}},
  \bibnamefont{and} \bibinfo{author}{\bibfnamefont{G.~A.}
  \bibnamefont{Sawatzky}}, \bibinfo{journal}{Phys. Rev. Lett.}
  \textbf{\bibinfo{volume}{68}}, \bibinfo{pages}{3924} (\bibinfo{year}{1992}).

\bibitem[{\citenamefont{Sivan and Imry}(1986)}]{imry86}
\bibinfo{author}{\bibfnamefont{U.}~\bibnamefont{Sivan}} \bibnamefont{and}
  \bibinfo{author}{\bibfnamefont{Y.}~\bibnamefont{Imry}},
  \bibinfo{journal}{Phys. Rev. B} \textbf{\bibinfo{volume}{33}},
  \bibinfo{pages}{551} (\bibinfo{year}{1986}),
  \urlprefix\url{http://link.aps.org/doi/10.1103/PhysRevB.33.551}.

\bibitem[{\citenamefont{Ke et~al.}(2009)\citenamefont{Ke, Yang, Cortarolo, and
  Baranger}}]{ke09}
\bibinfo{author}{\bibfnamefont{S.~H.} \bibnamefont{Ke}},
  \bibinfo{author}{\bibfnamefont{M.}~\bibnamefont{Yang}},
  \bibinfo{author}{\bibfnamefont{S.}~\bibnamefont{Cortarolo}},
  \bibnamefont{and} \bibinfo{author}{\bibfnamefont{H.~U.}
  \bibnamefont{Baranger}}, \bibinfo{journal}{Nano Lett.}
  \textbf{\bibinfo{volume}{9}}, \bibinfo{pages}{1011} (\bibinfo{year}{2009}).

\bibitem[{\citenamefont{Widawsky et~al.}(2012)\citenamefont{Widawsky, Darancet,
  Neaton, and Venkataraman}}]{widawsky12}
\bibinfo{author}{\bibfnamefont{J.~R.} \bibnamefont{Widawsky}},
  \bibinfo{author}{\bibfnamefont{P.}~\bibnamefont{Darancet}},
  \bibinfo{author}{\bibfnamefont{J.~B.} \bibnamefont{Neaton}},
  \bibnamefont{and}
  \bibinfo{author}{\bibfnamefont{L.}~\bibnamefont{Venkataraman}},
  \bibinfo{journal}{Nano Lett.} \textbf{\bibinfo{volume}{354}},
  \bibinfo{pages}{12} (\bibinfo{year}{2012}).

\bibitem[{\citenamefont{Paulsson and Datta}(2003)}]{paulsson03}
\bibinfo{author}{\bibfnamefont{M.}~\bibnamefont{Paulsson}} \bibnamefont{and}
  \bibinfo{author}{\bibfnamefont{S.}~\bibnamefont{Datta}},
  \bibinfo{journal}{Phys. Rev. B} \textbf{\bibinfo{volume}{67}},
  \bibinfo{pages}{241403} (\bibinfo{year}{2003}),
  \urlprefix\url{http://link.aps.org/doi/10.1103/PhysRevB.67.241403}.

\bibitem[{\citenamefont{Segal}(2005)}]{segal05}
\bibinfo{author}{\bibfnamefont{D.}~\bibnamefont{Segal}},
  \bibinfo{journal}{Phys. Rev. B} \textbf{\bibinfo{volume}{72}},
  \bibinfo{pages}{165426} (\bibinfo{year}{2005}),
  \urlprefix\url{http://link.aps.org/doi/10.1103/PhysRevB.72.165426}.

\bibitem[{\citenamefont{Ke et~al.}(2008)\citenamefont{Ke, Yang, and
  Baranger}}]{ke08}
\bibinfo{author}{\bibfnamefont{S.-H.} \bibnamefont{Ke}},
  \bibinfo{author}{\bibfnamefont{W.}~\bibnamefont{Yang}}, \bibnamefont{and}
  \bibinfo{author}{\bibfnamefont{H.~U.} \bibnamefont{Baranger}},
  \bibinfo{journal}{Nano Lett.} \textbf{\bibinfo{volume}{8}},
  \bibinfo{pages}{3257} (\bibinfo{year}{2008}).

\bibitem[{\citenamefont{Cardamone et~al.}(2006)\citenamefont{Cardamone,
  Stafford, and Mazumbdar}}]{cardamone06}
\bibinfo{author}{\bibfnamefont{D.~M.} \bibnamefont{Cardamone}},
  \bibinfo{author}{\bibfnamefont{C.~A.} \bibnamefont{Stafford}},
  \bibnamefont{and}
  \bibinfo{author}{\bibfnamefont{S.}~\bibnamefont{Mazumbdar}},
  \bibinfo{journal}{Nano Lett.} \textbf{\bibinfo{volume}{6}},
  \bibinfo{pages}{2423} (\bibinfo{year}{2006}).

\bibitem[{\citenamefont{Bergfield et~al.}(2012)\citenamefont{Bergfield, Liu,
  Burke, and Stafford}}]{bergfield11}
\bibinfo{author}{\bibfnamefont{J.~P.} \bibnamefont{Bergfield}},
  \bibinfo{author}{\bibfnamefont{Z.-F.} \bibnamefont{Liu}},
  \bibinfo{author}{\bibfnamefont{K.}~\bibnamefont{Burke}}, \bibnamefont{and}
  \bibinfo{author}{\bibfnamefont{C.~A.} \bibnamefont{Stafford}},
  \bibinfo{journal}{Phys. Rev. Lett.} \textbf{\bibinfo{volume}{108}},
  \bibinfo{pages}{066801} (\bibinfo{year}{2012}).

\bibitem[{\citenamefont{Stefanucci and Kurth}(2011)}]{stefanucci11}
\bibinfo{author}{\bibfnamefont{G.}~\bibnamefont{Stefanucci}} \bibnamefont{and}
  \bibinfo{author}{\bibfnamefont{S.}~\bibnamefont{Kurth}},
  \bibinfo{journal}{Phys. Rev. Lett.} \textbf{\bibinfo{volume}{107}},
  \bibinfo{pages}{216401} (\bibinfo{year}{2011}).

\bibitem[{\citenamefont{Schmitteckert and Evers}(2011)}]{schmitteckert11}
\bibinfo{author}{\bibfnamefont{P.}~\bibnamefont{Schmitteckert}}
  \bibnamefont{and} \bibinfo{author}{\bibfnamefont{F.}~\bibnamefont{Evers}},
  \bibinfo{journal}{Phys. Chem. Chem. Phys.} \textbf{\bibinfo{volume}{13}},
  \bibinfo{pages}{14417} (\bibinfo{year}{2011}).

\bibitem[{\citenamefont{Schmitteckert and Evers}(2008)}]{schmitteckert08}
\bibinfo{author}{\bibfnamefont{P.}~\bibnamefont{Schmitteckert}}
  \bibnamefont{and} \bibinfo{author}{\bibfnamefont{F.}~\bibnamefont{Evers}},
  \bibinfo{journal}{Phys. Rev. Lett.} \textbf{\bibinfo{volume}{100}},
  \bibinfo{pages}{086401} (\bibinfo{year}{2008}).

\bibitem[{\citenamefont{Grimme}(2006)}]{Grimme2006}
\bibinfo{author}{\bibfnamefont{S.}~\bibnamefont{Grimme}},
  \bibinfo{journal}{Journal of Computational Chemistry}
  \textbf{\bibinfo{volume}{27}}, \bibinfo{pages}{1787} (\bibinfo{year}{2006}).

\bibitem[{\citenamefont{Peter}(2009)}]{MatthiasPeter}
\bibinfo{author}{\bibfnamefont{M.}~\bibnamefont{Peter}}, Master's thesis,
  \bibinfo{school}{Universit{\"a}t Karlsruhe} (\bibinfo{year}{2009}).

\bibitem[{\citenamefont{Meir and Wingreen}(1992)}]{meir92}
\bibinfo{author}{\bibfnamefont{Y.}~\bibnamefont{Meir}} \bibnamefont{and}
  \bibinfo{author}{\bibfnamefont{N.}~\bibnamefont{Wingreen}},
  \bibinfo{journal}{Phys. Rev. Lett.} \textbf{\bibinfo{volume}{68}},
  \bibinfo{pages}{2512} (\bibinfo{year}{1992}).

\bibitem[{\citenamefont{Solomon
  et~al.}(2008{\natexlab{a}})\citenamefont{Solomon, Andrews, Hansen, Goldsmith,
  Wasielewski, Van~Duyne, and Ratner}}]{solomon08a}
\bibinfo{author}{\bibfnamefont{G.~C.} \bibnamefont{Solomon}},
  \bibinfo{author}{\bibfnamefont{D.~Q.} \bibnamefont{Andrews}},
  \bibinfo{author}{\bibfnamefont{T.}~\bibnamefont{Hansen}},
  \bibinfo{author}{\bibfnamefont{R.~H.} \bibnamefont{Goldsmith}},
  \bibinfo{author}{\bibfnamefont{M.~R.} \bibnamefont{Wasielewski}},
  \bibinfo{author}{\bibfnamefont{R.~P.} \bibnamefont{Van~Duyne}},
  \bibnamefont{and} \bibinfo{author}{\bibfnamefont{M.~A.}
  \bibnamefont{Ratner}}, \bibinfo{journal}{J. Chem. Phys.}
  \textbf{\bibinfo{volume}{129}}, \bibinfo{pages}{054701}
  (\bibinfo{year}{2008}{\natexlab{a}}).

\bibitem[{\citenamefont{Solomon
  et~al.}(2008{\natexlab{b}})\citenamefont{Solomon, Andrews, Goldsmith, Hansen,
  Wasielewski, Van~Duyne, and Ratner}}]{solomon08b}
\bibinfo{author}{\bibfnamefont{G.~C.} \bibnamefont{Solomon}},
  \bibinfo{author}{\bibfnamefont{D.~Q.} \bibnamefont{Andrews}},
  \bibinfo{author}{\bibfnamefont{R.~H.} \bibnamefont{Goldsmith}},
  \bibinfo{author}{\bibfnamefont{T.}~\bibnamefont{Hansen}},
  \bibinfo{author}{\bibfnamefont{M.~R.} \bibnamefont{Wasielewski}},
  \bibinfo{author}{\bibfnamefont{R.~P.} \bibnamefont{Van~Duyne}},
  \bibnamefont{and} \bibinfo{author}{\bibfnamefont{M.~A.}
  \bibnamefont{Ratner}}, \bibinfo{journal}{J. Chem. Phys.}
  \textbf{\bibinfo{volume}{130}}, \bibinfo{pages}{17301}
  (\bibinfo{year}{2008}{\natexlab{b}}).

\bibitem[{\citenamefont{Guedon et~al.}(2012)\citenamefont{Guedon, Valkenier,
  Markussen, Thygesen, Hummelen, and van~der Molen}}]{guedon12}
\bibinfo{author}{\bibfnamefont{C.~M.} \bibnamefont{Guedon}},
  \bibinfo{author}{\bibfnamefont{H.}~\bibnamefont{Valkenier}},
  \bibinfo{author}{\bibfnamefont{T.}~\bibnamefont{Markussen}},
  \bibinfo{author}{\bibfnamefont{K.~S.} \bibnamefont{Thygesen}},
  \bibinfo{author}{\bibfnamefont{J.~C.} \bibnamefont{Hummelen}},
  \bibnamefont{and} \bibinfo{author}{\bibfnamefont{S.~J.} \bibnamefont{van~der
  Molen}}, \bibinfo{journal}{Nat. Nano.} \textbf{\bibinfo{volume}{7}},
  \bibinfo{pages}{305} (\bibinfo{year}{2012}).

\end{thebibliography}

%%%%%%%%%%%%%%%%%%%%%%%%%%%%%%%%%%%%%%%%%%%%%%%%%%%%%%%%%%%%%%%%%
\end{document}